\documentclass[aps,rmp,preprint,amsmath,amssymb,graphicx,longbibliography,unsortedaddress]{revtex4-1}

\usepackage{graphicx}
\usepackage{amsmath, wasysym}
\usepackage[dvipsnames]{xcolor}
\usepackage{url, hyperref}
\hypersetup{colorlinks = true, citecolor = MidnightBlue, urlcolor = cyan, linkcolor = MidnightBlue}

\begin{document}

\title{Quantum and Nonlinear Optics in Strongly Interacting Atomic Ensembles}

\author{C. R. Murray}
\author{T. Pohl}
\address{Max Planck Institute for the Physics of Complex Systems, Dresden, Germany}

\begin{abstract}
Coupling light to ensembles of strongly interacting particles has emerged as a promising
route toward achieving few photon nonlinearities. One specific way to implement
this kind of nonlinearity is to interface light with highly excited atomic Rydberg states by
means of electromagnetically induced transparency, an approach which allows freely
propagating photons to acquire synthetic interactions of hitherto unprecedented
strength. Here, we present an overview of this rapidly developing field, from classical
effects to quantum manifestations of the nonlocal nonlinearities emerging in such systems.
With an emphasis on underlying theoretical concepts, we describe the many
experimental breakthroughs so far demonstrated and discuss potential applications
looming on the horizon.
\end{abstract}

\maketitle

\tableofcontents{}

\newpage

\section{Introduction}

Photons do not interact with one another in vacuum. From a practical perspective, this defining property of light can be considered a very useful
one. In particular, it underlies the success of modern-day optical communication technologies, as it ultimately enables the low-loss transmission of
densely encoded information over large distances. However, there are many other emerging concepts in both fundamental and applied science that would greatly benefit from a controllable interaction between photons, either on a classical \cite{Boyd} or on a quantum \cite{OBrien2009,Chang2014} level. The endeavor to capitalize on these prospects has since catalyzed a new era of research into nonlinear optics at ever-decreasing light intensities.

\begin{figure}[!b]
\begin{center}
\includegraphics[width=0.8\textwidth]{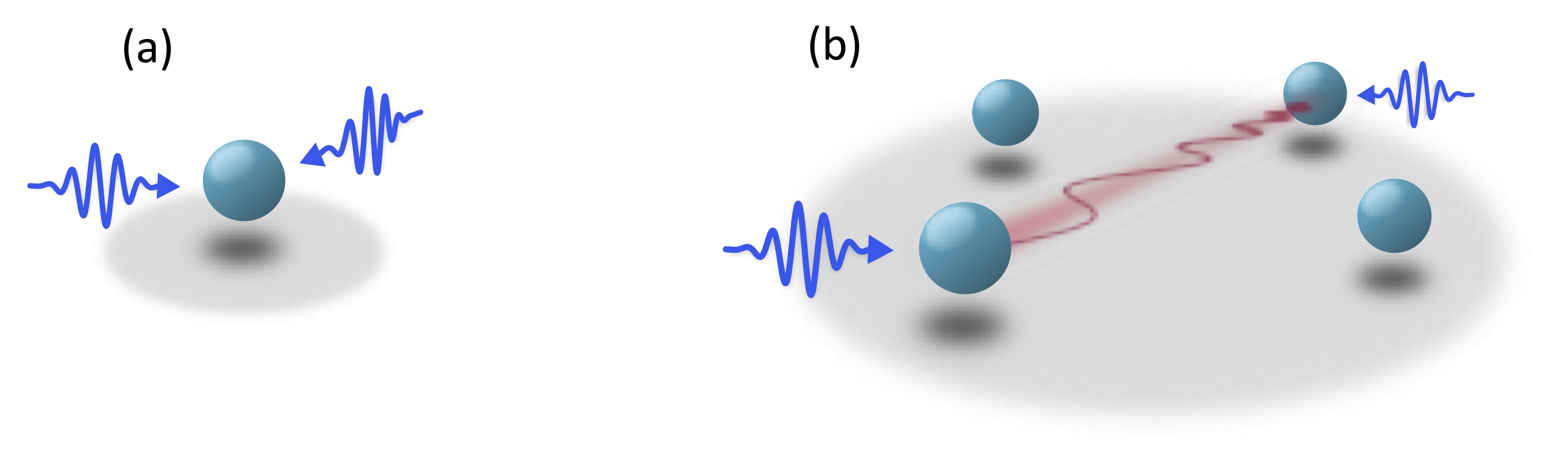}
\end{center}
\caption{\label{fig: Local to nonlocal}(a) Photons can interact via a local Kerr nonlinearity when they are coupled to a common scatterer. (b) While this interaction is diminished for light coupling to an ensemble, strong and nonlocal nonlinearities can arise from finite-range state-dependent interactions between particles.}
\end{figure}

Despite the fact that photons do not interact directly in vacuum, a strong synthetic photon-photon interaction can be mediated via their interaction
with an auxiliary material system. In order to realize a strong emergent photon-photon interaction, the medium in question must fulfill two major
requirements. First, the associated light-matter coupling must be strong and coherent so as to minimize photon loss, and second, the medium must feature such a strong optical nonlinearity that a single photon is sufficient to strongly modify its optical response.

The most explored approach is based on coupling photons to single quantum scatterers FIG.\ref{fig: Local to nonlocal}a, such as atoms \cite{Haroche2013}, molecules \cite{Wrigge2007}, or quantum dots \cite{Lodahl2015}. The strong nonlinearity available under such circumstances can be accredited to the saturable light absorption of a two-level system; as soon as it absorbs one photon, it can no longer absorb a second. The challenge here arises from the weakness of typical light-matter interactions, which renders the associated nonlinearity rather inefficient.

To understand why this is the case, consider a light pulse, with a transverse beam waist $\omega_0$, interacting with a single atom. The absorption cross-section $\sigma$ is maximal when the photon's frequency is resonant with the atomic transition frequency and, in this case, becomes proportional to the square of the photon's wavelength $\lambda$. Since diffraction typically limits the beam waist focussing to $\omega_0\gg\lambda$, the probability, $p\approx\lambda^2/\omega_0^2$, for single-photon scattering becomes very small, and with it, the strength of the optical nonlinearity.

\begin{figure*}[!b]
\begin{center}
\includegraphics[width=\textwidth]{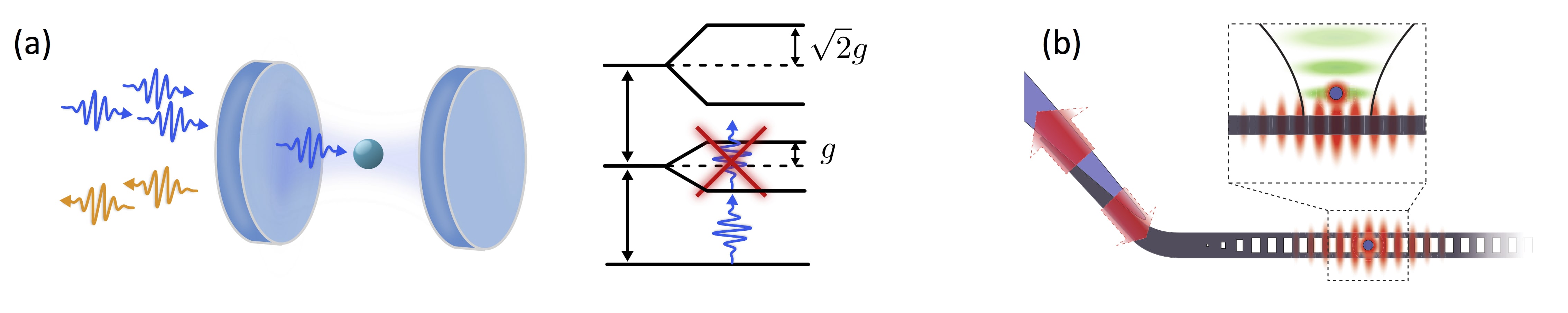}
\end{center}
\caption{\label{fig: Local Kerr nonlinearities} Examples of nonlinear optics settings based on strong light-matter coupling involving single scatterers. (a) Coupling photons to a single atom in an optical cavity can lead to a strong anharmonicity of the cavity occupation spectrum and induce a photon blockade that prevents multiple photons from entering the resonator in the strong coupling limit \cite{Birnbaum2005}. (b) Tight-mode confinement in nanophotonic structures can also reach this regime by trapping single atoms close to their surface to couple strongly to the evanescent field of a guided mode (see inset). Panel (b) adapted from \cite{Tiecke2014}.}
\end{figure*}

In order to resolve this issue, it is necessary to enhance the light-matter interaction strength beyond what is possible in the free space. One approach
to achieve this regime of strong coupling is to confine light to very small mode volumes by means of high-Q optical resonators, containing a single atom \cite{Birnbaum2005, Gleyzes2007, Reiserer2015} FIG.\ref{fig: Local Kerr nonlinearities}a. A strong optical nonlinearity then arises for large atom-cavity coupling, $g$, due to the anharmonicity of the Jaynes-Cummings ladder of cavity occupation eigenstates that develops \cite{Jaynes1963}, which makes transmission through the cavity sensitive to photon number. More recently, complementary approaches to achieving the strong coupling regime have been explored with the use of nanostructured optical media FIG.\ref{fig: Local Kerr nonlinearities}b, such as photonic crystal cavities   \cite{Thompson2013, Tiecke2014} or tapered nanofibers \cite{Vetsch2010}. Here, light is tightly guided below the free-space diffraction limit, such that by placing a single atom within the evanescent field of the guided mode, one can achieve a strong light-matter coupling. Such fabricated structures offer additional geometric flexibility, holding appeal for potential implementations of photonic quantum networks. Considerable efforts are currently being directed toward exploring different architectures based on this general philosophy, and we refer the reader to \cite{Chang2014}, for a more exhaustive discussion of such recent developments.

An alternative approach to achieving an efficient light-matter interaction is rather to couple photons to particle ensembles \cite{Hammerer2010},
which in this case leverages on the collectively enhanced coupling of light to many identical quantum systems. This ultimately circumvents the need for
additional mode confinement of light and is naturally more robust to single-particle losses. Yet the very nature of the collective coupling to many particles in this case diminishes the intrinsic nonlinearity associated to a single saturable scatterer from before.

However, it is possible to reestablish a large optical nonlinearity in ensembles if the material constituents feature strong state-dependent interactions.
This provides a fundamentally distinct approach to quantum nonlinear optics in which two distant photons can interact via a nonlocal nonlinearity
mediated by long-ranged particle-particle interactions, rather than interacting locally in space by coupling to a common particle FIG.\ref{fig: Local to nonlocal}b.

While such conditions are difficult to meet in conventional materials, cold gases of Rydberg atoms offer the type of exaggerated interactions to
make this idea work. As pointed out in pioneering work by Friedler et al. \cite{Friedler2005}, the combination of electromagnetically induced transparency (EIT) and Rydberg-Rydberg atom interactions precisely meets the two above requirements of efficient low-loss light-matter coupling and strong emergent nonlinearities. The first experimental observation of EIT in a Rydberg gas \cite{Mohapatra2007} demonstrated the viability of this
approach and paved the way for subsequent measurements of highly nonlinear propagation of classical \cite{Pritchard2010} and quantum \cite{Peyronel2012} light fields, where nonlinear phenomena begin to manifest at the level of individual light quanta in the latter case. 

These seminal works have since ushered in rapidly increasing activities, both on the theoretical understanding of the underlying physics and on the
advancement of experimental approaches to harness this novel mechanism for achieving strong optical nonlinearities. Recent experimental breakthroughs are summarized in an excellent article by Firstenberg et al. \cite{Firstenberg2016}. Here we review developed theoretical concepts and discuss their connection to experimental observations and potential applications.

In the remainder of this section, we will give a brief introduction to the essential concepts underlying EIT and Rydberg physics that are relevant for content of this article. However, for a more exhaustive overview of these subjects, we refer the reader to Fleischhauer et al. \cite{Fleischhauer2005} and Saffman et al. \cite{Saffman2010} respectively.

\begin{figure}[!b]
\begin{center}
\includegraphics[width=0.9\textwidth]{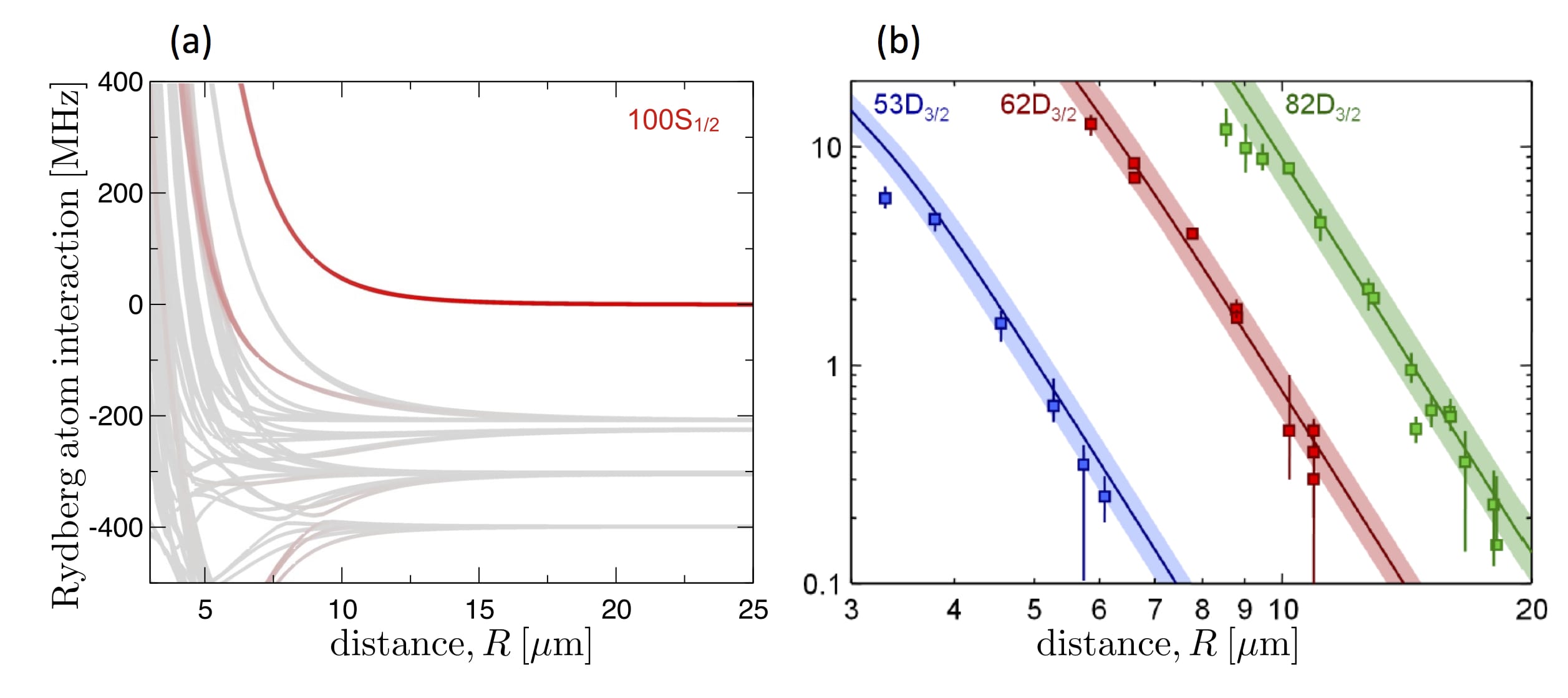}
\end{center}
\caption{\label{fig:vdW} (a) Rydberg-Rydberg atom interaction potentials around the ($100S_{1/2},100S_{1/2}$)-asymptote of Rubidium atoms. The red colouring indicates the two-photon coupling strength to a given potential curve. (b) Measured interactions between $nD_{3/2}$ Rydberg atoms for different principal quantum numbers, $n$, in comparison to theoretical calculations (lines). Panel (b) adapted from \cite{Beguin2013}}
\end{figure}

\subsection{Cold Rydberg gases}
\label{subsec:Cold Rydberg gases}
Rydberg atoms are atoms in highly excited electronic states with a large principal quantum number $n$. Due to the well-known $\sim n^2$ scaling of the electron's Bohr radius, the size of such atoms can become enormous, reaching micrometer scales for typical excitation levels realized in current
experiments. Such large length scales endow Rydberg atoms with a number of exaggerated properties that make them ideal candidates for the coherent manipulation of strongly interacting quantum systems \cite{Saffman2010}. In particular, the small spatial overlap between the electronic ground state wavefunction and the inflated Rydberg state implies a small spontaneous decay rate, ie, a very long lifetime that increases as $\sim n^3$ and can be on the order of $100 \mu s$ in experiments. At the same time, the weak Coulomb interaction between the nucleus and the Rydberg electron renders such Rydberg states very sensitive to their environment and, for instance, gives rise to an electric polarizability that scales as $\sim n^7$. This large polarizability has been exploited in \cite{Mohapatra2008} to realize a drastically enhanced dc-Kerr effect in a cold Rydberg gas. Perhaps the most striking consequence of this high sensitivity is a vast enhancement of the interaction between atoms which can exceed that of ground state atoms by more than $10$ orders of magnitude for typical Rydberg states produced in the laboratory. While the interaction can become rather complex at smaller interatomic distances FIG.\ref{fig:vdW}a, its asymptotic behavior is typically that of a simple van der Waals potential
\begin{equation}
V(R)=\frac{C_6}{R^6}\;.
\end{equation}
The van der Waals coefficient, $C_6\sim n^{11}$, exhibits an even stronger dependence on the principal quantum number and therefore becomes extremely large for high-lying Rydberg states. Unlike for the generically attractive van der Waals interactions between ground state atoms, the potential for Rydberg states can have either sign or feature an angular dependence, offering a high level of control over the interactions in such systems. The accuracy of this simple picture has been impressively demonstrated in a number of recent measurements \cite{Beguin2013,Barredo2014,Ravets2014,Ravets2015}, an example of which is shown in FIG.\ref{fig:vdW}b.

Due to the fragile nature of highly excited Rydberg atoms, systematic studies of such large interactions only became possible with the availability
of laser-cooled gases that permitted spectroscopic interrogations of effectively ``frozen Rydberg gases'' \cite{Anderson1998,Mourachko1998} on timescales on which atomic motion does not affect the laser-driven dynamics of internal atomic states. The most prominent phenomenon relevant
to such a scenario is the so-called Rydberg blockade effect \cite{Jaksch2000,Lukin2001}, which is a strong suppression of multiple Rydberg excitations due to the two-body level shifts arising from the strong interactions. Describing a dense atomic ensemble in a continuum representation, the underlying Rydberg-Rydberg interaction Hamiltonian which captures this physics can be written as
\begin{equation}\label{eq:VHam}
\hat{H}_I=\int {\rm d}{\bf r}{\rm d}{\bf r}^\prime \hat{S}^\dagger({\bf r})\hat{S}^\dagger({\bf r}^\prime) V({\bf r}-{\bf r}^\prime)\hat{S}({\bf r}^\prime)\hat{S}({\bf r})
\end{equation}
where $\hat{S}^\dagger({\bf r})$ is the Rydberg atom creation operator. The critical distance $z_{\rm b}$, below which the blockade is effective, can then be defined by the point at which the interaction-induced level shift exceeds the laser line width $\Gamma_{\rm exc}$ of the Rydberg-state excitation process, ie, $\Gamma_{\rm exc}=V(z_{\rm b})$, and can take on large values $z_{\rm b}\sim 10\mu$m.

Since its conception, the blockade effect has generated a wealth of promising ideas, e.g., for applications of Rydberg atom ensembles as quantum information processors \cite{Saffman2010,Keating2013} or quantum simulators \cite{Weimer2010,Glaetzle2015,VanBijnen2015}. In small atomic ensembles whose spatial extent is below $z_{\rm b}$, the excitation light can only couple the $N$-atom ground state to a single Rydberg excitation that is coherently shared between all atoms. The light-matter coupling to such a ``super atom'' is collectively enhanced by $\sqrt{N}$, and therefore appears ideally suited as a source of single photons \cite{Lukin2001,Saffman2002} or atoms \cite{Saffman2002,Ebert2014}, or to exploit optical nonlinearities of single quantum emitters \cite{Guerlin2010}, as described previously.

\begin{figure}[!t]
\begin{center}
\includegraphics[width=\textwidth]{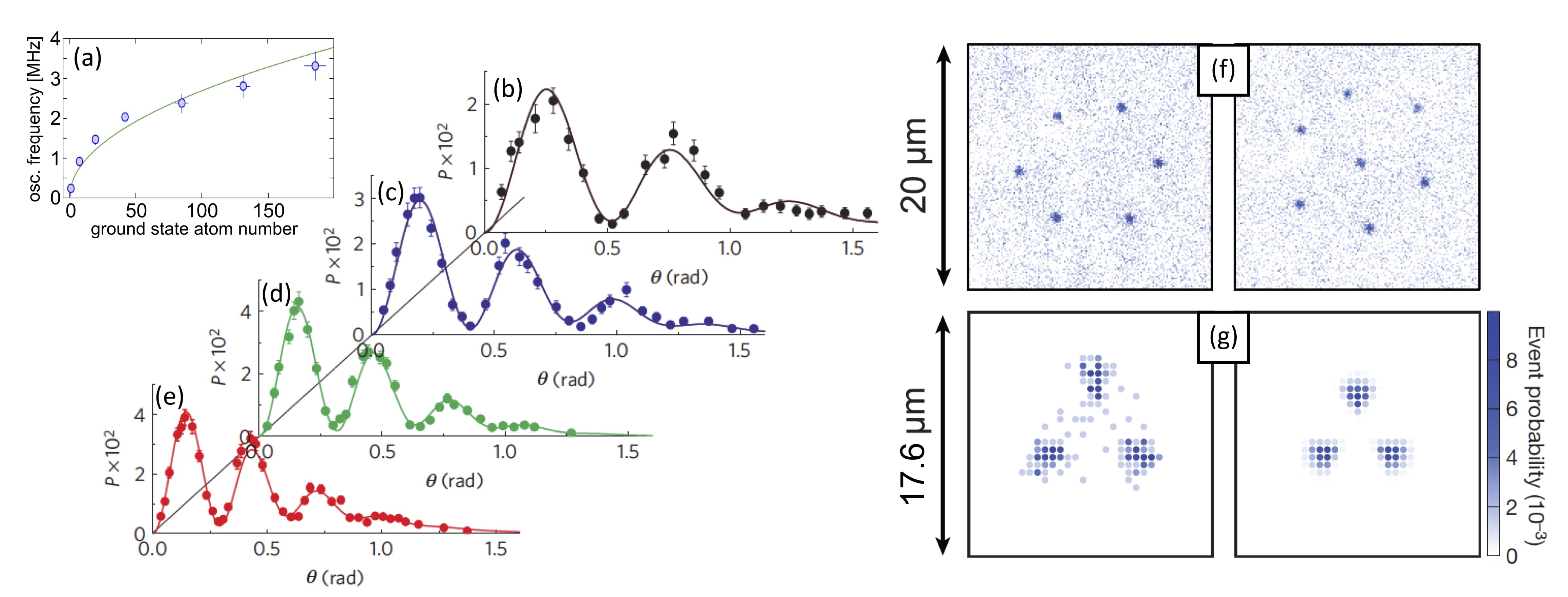}
\end{center}
\caption{\label{fig:blockexp} Different experimental demonstrations of the Rydberg blockade. (a) Direct detection of Rydberg atoms in an optical lattice reveals the expected $\sqrt{N}$-enhancement of the Rabi oscillation frequency for a fully blocked lattice of $N$ atoms \cite{Zeiher2015}. (b-e) Such an enhancement of Rabi oscillations has also been observed by mapping the excited Rydberg states onto photons \cite{Dudin2012a}. (f) Spatial imaging of Rydberg excitations in extended systems directly reveals the blockade mechanism which can produce ordered arrangements during the laser-excitation process \cite{Schauss2015}. (g) Averaging over many such projective measurements yields a Rydberg-density distribution corresponding to the $N$-body wave function of the system, which is shown for the $3$-excitation component and agrees well with theoretical calculations (left) \cite{Schauss2015}. Panels (a), (b-e), (f) and (g) adapted from \cite{Zeiher2015}, \cite{Dudin2012a}, \cite{Schauss2015} and \cite{Schauss2012}, respectively.}
\end{figure}

Experiments found evidence of the Rydberg blockade in extended ensembles of different atomic species and various types of interactions by observing a strong suppression of the fraction of laser-exited states \cite{Singer2004,Tong2004,Vogt2006,Heidemann2007}, or through modifications in the counting statistics of produced Rydberg atoms \cite{Liebisch2005}. A major step towards the coherent manipulation of Rydberg systems has been made with the development of optical detection techniques that are less invasive than previously used field-ionisation schemes. This includes a mapping of generated Rydberg excitations onto photons \cite{Dudin2012,Dudin2012a} and state-selective fluorescence imaging of ground state atoms in large-spacing micro-trap arrays \cite{Urban2009,Gaetan2009} or in optical lattices via quantum gas microscope techniques \cite{Schauss2012}. For example, this made it possible to observe the transition to ``super atom'' behaviour, as outlined above \cite{Ebert2015,Zeiher2015}, and enabled the direct imaging of the Rydberg blockade effect in extended many-body systems \cite{Schauss2012,Schauss2015} (see FIG.\ref{fig:blockexp}).

The key advance for the optical and nondestructive probing of extended frozen Rydberg gases has been made in \cite{Mohapatra2007}, using the narrow transmission resonances that emerge for laser-driven three-level systems under conditions of electromagnetically induced transparency (EIT). In addition to the strong interactions between Rydberg states, the phenomenon of EIT provides the second crucial ingredient for realizing optical nonlinearities in Rydberg gases, and will therefore be discussed briefly in the following section.

\subsection{Electromagnetically induced transparency with Rydberg atoms}
\begin{figure}[!b]
\begin{center}
\includegraphics[width=0.95\textwidth]{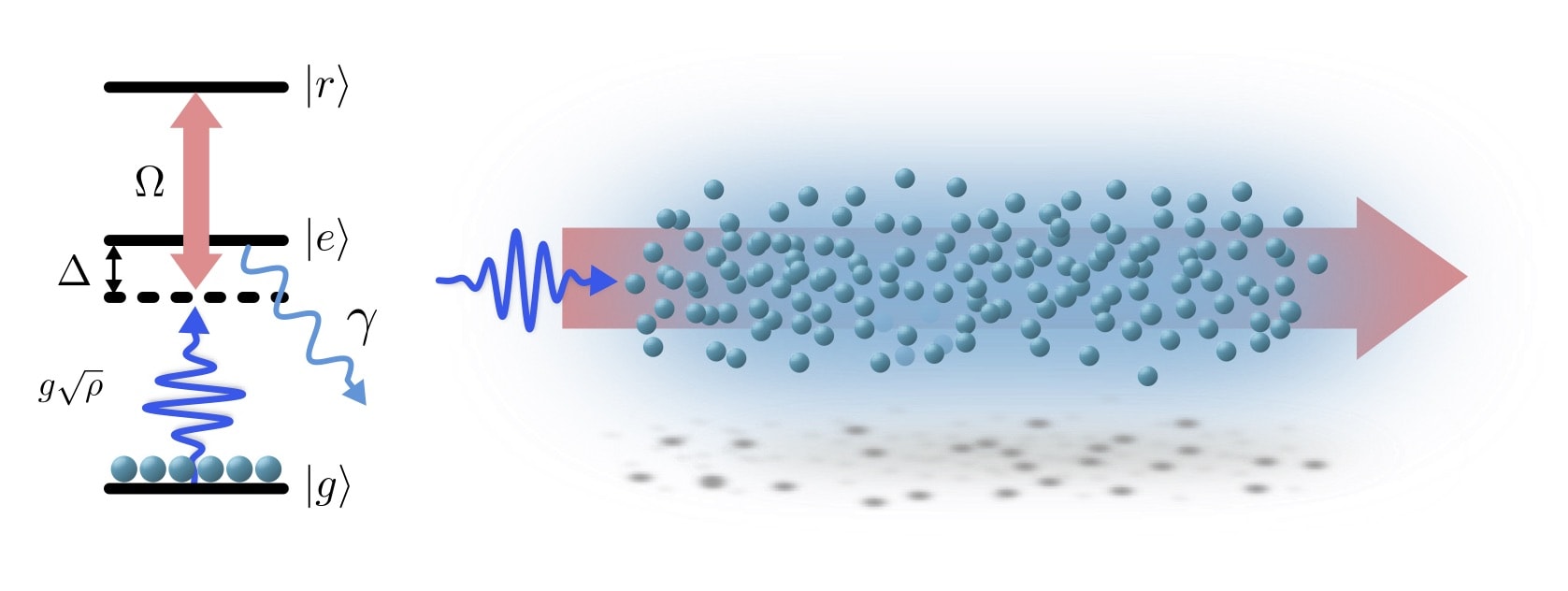}
\end{center}
\vspace{-5ex}
\caption{\label{fig: EIT setup} Photons are coupled to highly excited Rydberg states in a cold atomic ensemble under EIT conditions, in which they drive a low-lying transition between two states $|g\rangle$ and $|e\rangle$, whilst an auxiliary control field completes the resonant two-photon coupling to a Rydberg state $|r\rangle$. The probe field can be detuned by an amount $\Delta$ from the intermediate state $|e\rangle$, which is short-lived and decays with a rate $\gamma$.}
\end{figure}

EIT can arise in three-level atoms that are driven by two optical fields. In the present context, the relevant excitation scheme consists of a weak probe field which couples some ground state $|g\rangle$ to a low lying excited state $|e\rangle$, whilst a stronger control field then drives the transition from $|e\rangle$ to a high lying Rydberg state $|r\rangle$, and establishes a resonant two-photon coupling between the ground and Rydberg state (FIG.\ref{fig: EIT setup}). While EIT is more commonly considered for $\Lambda$-type level schemes involving two stable ground states, the long lifetime of the excited Rydberg state yields a virtually equivalent situation for the resulting ladder-type level structure.

In essence, EIT is an effect that exploits quantum interference to cancel the linear optical response (ie, refraction and absorption) of the probe transition when the control field is applied. This concept was originally discussed by \cite{Harris1990}, and the first demonstration of the effect was soon after realized by \cite{Boller1991}. However, the quantum interference underlying EIT is a rather sensitive effect that can be easily perturbed. In fact, the philosophy behind obtaining optical nonlinearities in a Rydberg-EIT setting is precisely based on exploiting this feature, ie, using the Rydberg blockade to destroy the transparency for more than a single photon.

A quantum treatment of light propagation under EIT conditions \cite{Fleischhauer2000,Fleischhauer2002} can be formulated in terms of the
slowly varying amplitude operator $\hat{\mathcal{E}}^{\dagger}(z)$ that describes the creation of probe photon at position $z$. It often suffices to treat the light propagation in a one-dimensional approximation and consider only the dynamics along the propagation direction of the field. In terms of characterizing the atomic system, the limit of weak probe-field intensities justifies a linearized treatment of the atomic state dynamics \cite{Fleischhauer2000,Fleischhauer2002}, which is then captured solely by the Bosonic operators
\begin{align}
\hat{P}(z) & = \frac{1}{\sqrt{\rho}} \sum_i \delta(z - z_i) |g_i \rangle \langle e_i | \\
\hat{S}(z) & = \frac{1}{\sqrt{\rho}} \sum_i \delta(z - z_i) |g_i \rangle \langle r_i | 
\end{align}
where the sum is over all atoms at random positions $z_i$ in the medium and $\rho=\sum_i \delta(z - z_i)$ denotes the density of the atomic gas. The field operator $\hat{P}^{\dagger}(z)$ creates a collective polarization coherence (on the $|g\rangle - |e\rangle$ transition) whilst $\hat{S}^{\dagger}(z)$ creates a collective Rydberg spin wave coherence (on the $|g\rangle - |r\rangle$ transition). The propagation dynamics of $\hat{\mathcal{E}}(z)$ under the interaction with these two matter fields is then governed by the Heisenberg equations 
\begin{align}
\label{eq:HB1}
\partial_t \hat{\mathcal{E}}(z) & = -c \partial_z  \hat{\mathcal{E}}(z) + i g\sqrt{\rho} \hat{P}(z) \\
\label{eq:HB2}
\partial_t \hat{P}(z) & = i g\sqrt{\rho} \hat{\mathcal{E}}(z)  + i \Omega \hat{S}(z) + i \left[ \Delta + i \gamma \right] \hat{P}(z) \\
\label{eq:HB3}
\partial_t \hat{S}(z) & = i \Omega \hat{P}(z) - i \int dz^{\prime} V(z - z^{\prime}) \hat{S}^{\dagger}(z^{\prime})\hat{S}(z^{\prime}) \hat{S}(z) 
\end{align}
where $c$ is the vacuum speed of light, $g\sqrt{\rho}$ is the collectively enhanced coupling between $\hat{\mathcal{E}}(z)$ and $\hat{P}(z)$ (where $g$ is the single atom coupling strength), and $2\Omega$ is the Rabi frequency of the classical control field coupling $\hat{P}(z)$ and $\hat{S}(z)$. Since the state $|e\rangle$ is typically short-lived, the polarisation $\hat{P}(z)$ decays with a rate $\gamma$ that defines the natural line width of the $|g\rangle - |e\rangle$ transition. On the other hand, the small decay rate of the Rydberg state can be neglected under most conditions. As will be repeatedly stressed, the single-photon detuning $\Delta$ from the intermediate state plays an important role as it offers tunability in the dissipative contribution to the optical response of the medium stemming from the Rydberg-Rydberg interactions, $V(z)$. The interaction term in Eq.(\ref{eq:HB3}) directly follows from the interaction Hamiltonian introduced in the previous section in Eq.(\ref{eq:VHam}), and describes the mutual level shifts $V(z-z^{\prime})$ experienced by proximate Rydberg excitations. 

\begin{figure}[!t]
\begin{center}
\includegraphics[width=0.75\textwidth]{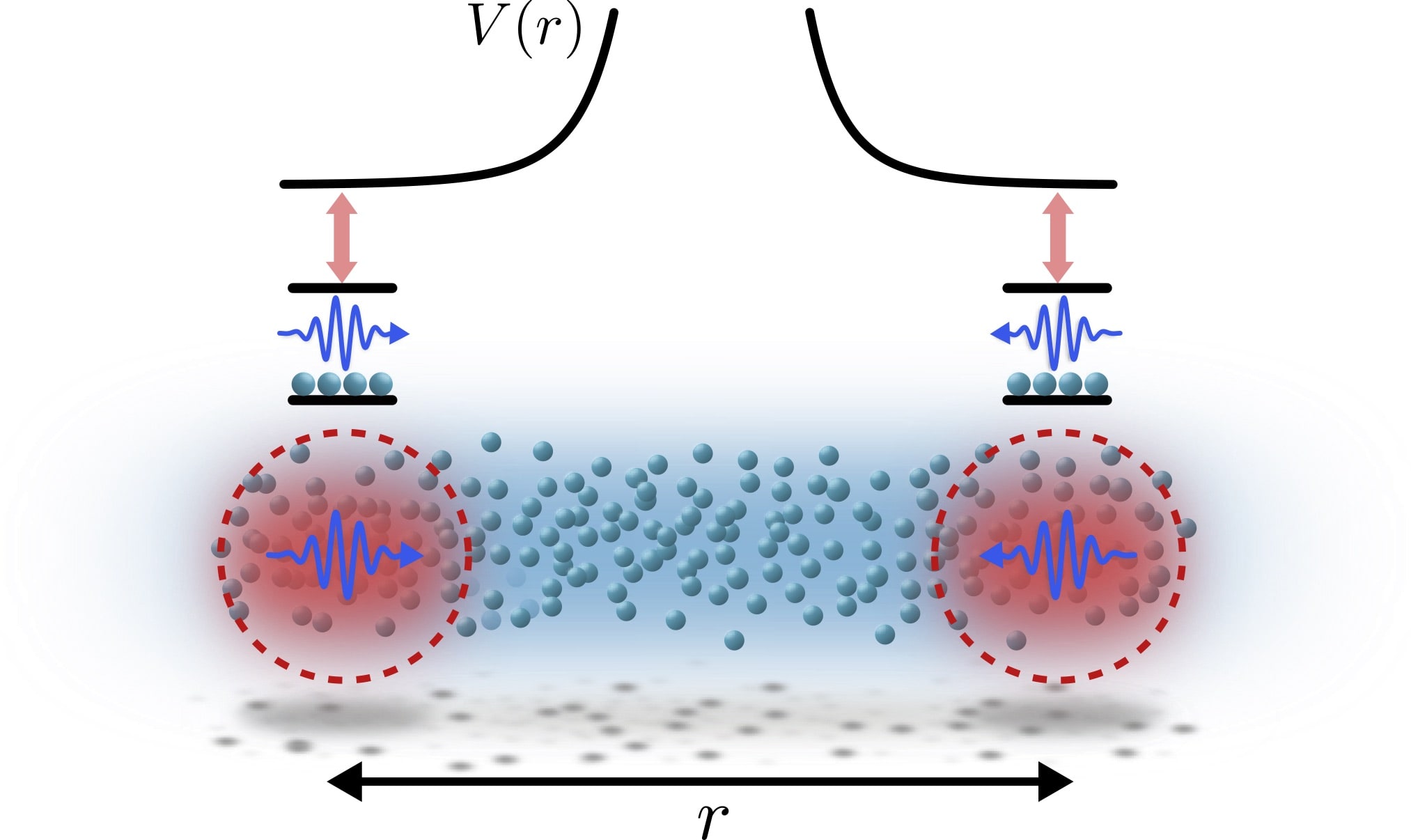}
\end{center}
\caption{\label{fig: Rydberg EIT} Two photons at large separation $r$ from one another can propagate transparently as Rydberg polaritons under EIT conditions. However, as they approach one another, the Rydberg state of one polariton will experience a spatially dependent shift $V(r)$ due to the Rydberg interactions which breaks EIT conditions. This exposes an absorbing two-level medium of spatial extent $\sim 2z_b$, where $z_b$ denotes the blockade radius.}
\end{figure}

In the absence of the control beam, the probe field simply interacts with the bare two-level system formed by $|g\rangle$ and $|e\rangle$, which on resonance, $\Delta=0$, causes significant photon loss due to rapid spontaneous emission from the state $|e\rangle$. The probe field amplitude thus suffers an exponential attenuation $e^{-OD}$ after having traversed a length $L$ of the medium, where
\begin{equation}
OD = \frac{g^2 L \rho}{c \gamma}
\end{equation}
denotes the optical depth of the medium that effectively characterises the collective optical coupling strength to the ensemble atoms, as reflected by the linear scaling with the atomic density $\rho$. The two-photon resonant coupling of the control field, $\Omega$, however, induces a spectral transparency window, in which absorption is suppressed over a frequency width $\Gamma_{\text{EIT}} = \Omega^2/|\Delta + i \gamma|$ \cite{Fleischhauer2005}. So if the probe light features a sufficiently narrow spectral width with respect to $\Gamma_{\text{EIT}}$, it can propagate with little loss. 

Additionally, the dispersive properties of the medium are also profoundly altered by the applied control field, the primary effect of which is a drastic reduction in the probe field's group velocity by a factor $\sim\Omega^2/(g^2\rho)$ \cite{Harris1992}. The dramatic consequences of this effect where highlighted in experiments by \cite{Hau1999}, where light was slowed to a remarkably low group velocity of $17{\rm m}/{\rm s}$. 
A dynamic tunability in the group velocity, offered by a time-varying control field, also permits the stoppage of light altogether, and protocols have been developed \cite{Fleischhauer2000,Fleischhauer2002,Gorshkov2007} and implemented \cite{Liu2001, Novikova2007} to achieve such light storage. This provides a particularly powerful approach to photonic quantum memories \cite{Lukin2003, Chaneliere2005} and will be central to the applications discussed in Section \ref{sec: All-optical switch} and \ref{sec: Two-photon phase gate}.

\begin{figure}[!b]
\begin{center}
\includegraphics[width=0.85\textwidth]{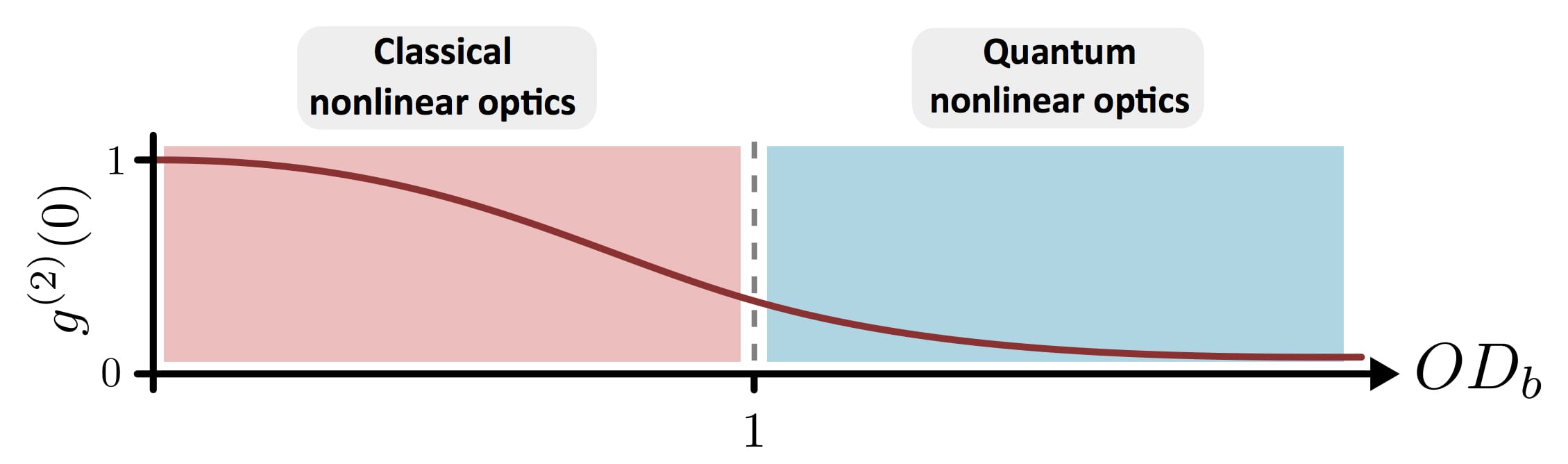}
\end{center}
\caption{\label{fig: g2 density} The character of the optical nonlinearity in a Rydberg EIT setting is crucially determined by the optical depth per blockade radius $OD_b$. At low low values $OD_b\ll1$, interactions do not affect the photon statistics of a coherent input field ($g^{(2)}(0)=0$) but can give rise to strong classically nonlinear effects. When $OD_b>1$ the emerging optical nonlinearities can generate nonclassical states of light, and affect its propagation on the level of single photons.}
\end{figure}

From a different perspective, EIT can be accredited to the formation of a so-called dark-state polariton \cite{Fleischhauer2000,Fleischhauer2002}
\begin{equation}
\label{eq:DSP}
\hat{\Psi}(z) = \frac{\Omega \hat{\mathcal{E}}(z) - g\sqrt{\rho} \hat{S}(z)}{\sqrt{\Omega^2+g^2\rho}}
\end{equation}
that corresponds to a quasi-particle comprised of electromagnetic and collective atomic excitations and can propagate in a near lossless and from-stable fashion. Importantly though, the polariton inherits the characteristics from both of its constituents: it gains kinetics from its photonic component, allowing it to propagate, and gains interactions from its Rydberg spin wave component, allowing it to interact with other polaritons. This polariton-polariton interaction then maps onto an effective photon-photon interaction due to the dipole blockade, which ultimately prevents two photons from simultaneously propagating transparently if they are too closely separated due to the breaking of EIT conditions, FIG.\ref{fig: Rydberg EIT}. The characteristic range of this interaction is then defined as the spatial separation $z_b$ at which the Rydberg atom van der Waals potential exceeds the EIT linewidth, i.e., $V(z_b) = \Gamma_{\text{EIT}}$. With typical EIT linewidths on the order of a few MHz, the enormous interactions between Rydberg atoms discussed in Section \ref{subsec:Cold Rydberg gases} (see FIG.\ref{fig:vdW}), thus, suggest large interaction ranges in excess of $\sim10\mu$m, well beyond the optical wave length of the probe photons.

The back-action of the Rydberg interactions on the probe field can be characterised in terms of the optical depth per blockade radius, $OD_b=g^2\rho z_b/(c\gamma)$, exposed by the interaction blockade. In this way, the strength of the emergent nonlinearity is intimately related to $OD_b$, and it ultimately defines the transition between domains of classical ($OD_b\ll1$) and quantum ($OD_b>1$) nonlinear optics. In particular, nonclassical correlations can be characterised by an equal-time two-photon correlation function $g^{(2)}(0)\neq1$ (FIG.\ref{fig: g2 density}). The remainder of the article is structured according these two distinct regimes, beginning with a discussion of the classical domain in the following section.

\section{Classical nonlinear optics}
\label{sec:Classical nonlinear optics}
As discussed above, the nature of nonlinear photon propagation is critically determined by the value of the optical depth, $OD_{\rm b}$, per blockade volume. In this section, we will focus on the limit of small $OD_{\rm b}$, in which emerging photon correlations are weak and light propagation can be described in terms of coherent photon fields. As we will see, discarding photon-photon and photon-atom correlations permits to develop an intuitive understanding of interaction effects based on nonlinearities of the classical photon-field amplitude $\mathcal{E}=\langle\hat{\mathcal{E}}\rangle$ that emerge from strong correlations between the interacting Rydberg excitations.

\subsection{Optical nonlinearities}
\label{subsec:Optical nonlinearities}
\begin{figure}[!t]
\begin{center}
\includegraphics[width=0.87\textwidth]{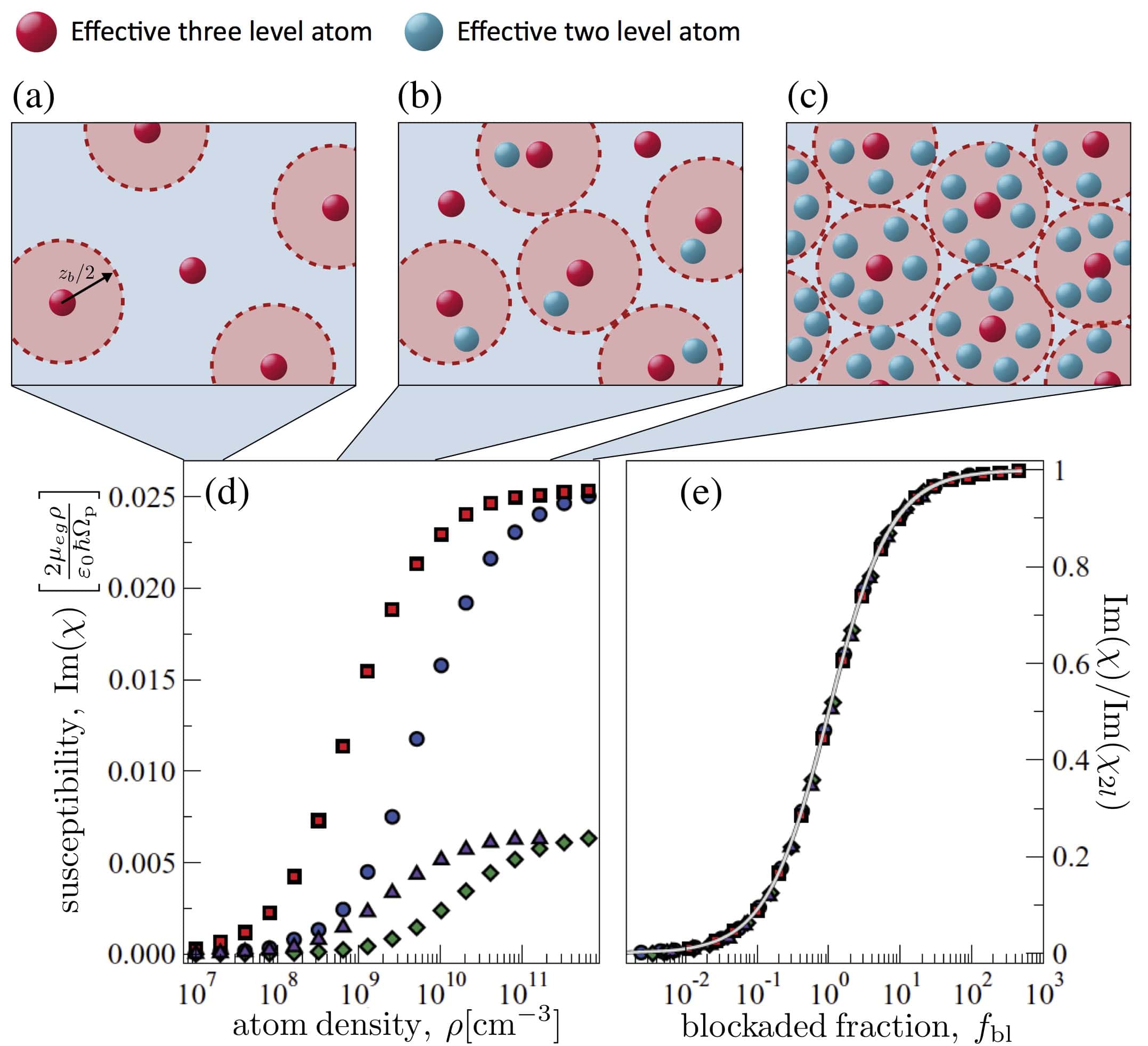}
\end{center}
\caption{\label{fig:MCresults} (a-c) Illustration of the basic mechanism leading to a density-dependent optical response shown in (d). The displayed Monte Carlo results are obtained for a cold rubidium gas excited to $nS_{1/2}$ Rydberg states with $\Omega=2$MHz and $\Omega_{\rm p}=1$MHz, $n=70$ (squares); $\Omega_{\rm p}=1$MHz, $n=50$ (circles); $\Omega_{\rm p}=0.5$MHz, $n=70$ (triangles); $\Omega_{\rm p}=0.5$MHz, $n=50$ (diamonds). Upon scaling by the two-level response all data points collapse onto a universal curve as a function of the blockaded fraction $f_{\rm bl}$. Panels (d-e) are reproduced from \cite{Ates2011}.}
\end{figure}
In order to develop an intuitive picture of the basic mechanism behind the emerging photon interactions, let us first consider the most simple scenario in which the probe field is assumed to be a static quantity that determines the Rabi frequency $\Omega_{\rm p}$ for the optical coupling of the lower $|g\rangle\leftrightarrow|e\rangle$ transition in FIG.\ref{fig: EIT setup}. This leaves us with a random ensemble of $i=1,...,N$ three-level atoms that are coherently driven by the two laser fields $\Omega_{\rm p}$ and $\Omega$ and subject to dissipation due to the intermediate state decay with a rate $\gamma$ (see FIG.\ref{fig: EIT setup}). Analogously to Eq.(\ref{eq:DSP}), the single-atom steady state corresponds to a dark state \cite{Fleischhauer2005}
\begin{equation}\label{eq:darkstate}
|D_i\rangle=\frac{\Omega_{\rm p}|r_i\rangle-\Omega|g_i\rangle}{\sqrt{\Omega_{\rm p}^2+\Omega^2}}
\end{equation}
that does not couple to the laser fields and is immune to decay from the intermediate state $|e_i\rangle$. The strong mutual Rydberg-state interactions, however, profoundly modify this simple picture and lead to subtle few-body effects that where found to promote nonclassical light emission from interacting atom pairs \cite{Pritchard2012b,Xu2015}, and shown to enable dissipative preparation of entangled $N$-atom states \cite{Petrosyan2013,Carr2013,Rao2014,Cano2014} as well as quantum gate operations \cite{Moller2008,Muller2009a,Rao2013,PhysRevA.89.030301}.

The theoretical description of large atomic ensembles presents a formidable task that remains beyond the reach of present computational capabilities, and has thus motivated the development of suitable approximations to this problem. In addition to recently proposed variational approaches \cite{Weimer2016,Weimer2015,Weimer2015a}, cluster expansions that approximate higher-order particle correlations have been successfully applied to describe experiments and understand interaction effects on coherent population trapping \cite{Schempp2010}.

Another, computationally powerful, approach exploits the strong dissipation of the intermediate state to approximately represent the many-body quantum system as a classical problem that only involves the atomic level populations, whose dynamics are then governed by transition rates \cite{Ates2007,Ates2007a,Ates2006}. Interaction effects are incorporated through the level shifts exerted by a produced Rydberg excitation on surrounding atoms which affects their transition rates and can thereby lead to strongly correlated excitation dynamics. The apparent advantage of this approach is that it still incorporates all $3^N$ $N$-body states of the original quantum problem, but permits an efficient solution via classical Monte Carlo sampling with remarkable accuracy, as thoroughly investigated in \cite{Schonleber2014}. Aside from applications to nonlinear optics, this method has been applied to address questions concerning the non-equilibrium physics of such driven-dissipative systems \cite{Petrosyan2013,Lesanovsky2014,Hoening2014,Sanders2014,Mattioli2015,Sanders2015}, and was used to model Rydberg gas experiments \cite{Ates2006,Schempp2014,Urvoy2015,Valado2016}. Importantly, such classical simulations still recover the dark state physics, Eq.(\ref{eq:darkstate}), arising from quantum interference, and allow one to obtain the optical susceptibility of the Rydberg-EIT medium
\begin{equation}
\chi=\frac{2\mu_{eg}^2\rho}{\varepsilon_0\hbar\Omega_{\rm p}}N^{-1}{\rm Tr}\left(\hat{\rho}_N\sum_i |g_i\rangle\langle e_i|\right)
\end{equation}
from the classically sampled $N$-body density matrix $\hat{\rho}_N$, as shown \cite{Ates2011}. Here, $\mu_{eg}$ denotes the dipole matrix element of the probe transition and $\varepsilon_0$ is the vacuum permittivity. The conceptual and numerical simplicity of this approach permits various extensions, for example, to account for higher order $n$-body excitation processes \cite{Heeg2012}, atomic motion \cite{Sibalic2016} or additional decoherence mechanisms \cite{Garttner2013}. 

The obtained susceptibilities indeed reveal a strong nonlinear dependence on the probe-field amplitude \cite{Ates2011,Heeg2012,Garttner2013,Garttner2014} that can be readily understood within the underlying classical picture, FIG.\ref{fig:MCresults}. In the absence of interactions each atom individually settles into the dark steady state Eq.(\ref{eq:darkstate}) such that ${\rm Tr}(\hat{\rho}_N |g_i\rangle\langle e_i|)=0$ and the optical response of the atomic medium vanishes. An excited Rydberg atom, however, imparts strong level shifts onto to the atoms in its vicinity and thereby turns a number $\sim z_{\rm b}^3\rho$ of atoms into two-level systems that now feature sizeable optical absorption
\begin{equation}
\chi_{2l}=\frac{2\mu_{eg}^2\rho}{\varepsilon_0\hbar\Omega_{\rm p}}N^{-1}{\rm Tr}\left(\hat{\rho}_N^{(2l)} |g_i\rangle\langle e_i|\right)\approx i\frac{\mu_{eg}^2\rho}{\varepsilon_0\hbar\gamma},
\end{equation}
where $\hat{\rho}_N^{(2l)}$ denotes the steady state density matrix of independent two-level atoms. For very small atomic densities, the relative contribution of such blockaded atoms is small such that the total optical susceptibility remains vanishingly small, $\chi\approx0$ (FIG.\ref{fig:MCresults}a). However, with increasing density, $\chi$ concomitantly increases and eventually approaches a finite value once the Rydberg excitation is saturated by the interaction blockade (FIG.\ref{fig:MCresults}b and c). In the high density limit the optical response is, hence, dominated by blockaded atoms such that the saturation value simply corresponds to the two-level susceptibility, $\chi_{2l}$, as shown in FIG.\ref{fig:MCresults}d.

\begin{figure}[!b]
\begin{center}
\includegraphics[width=\textwidth]{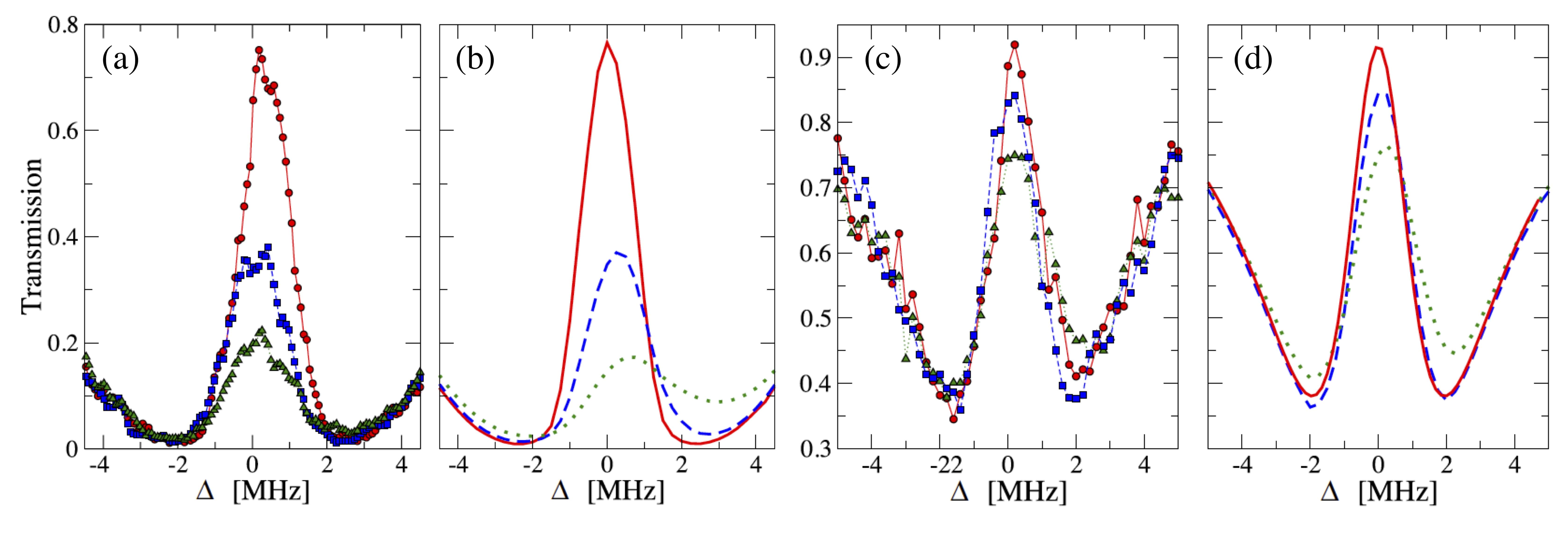}
\end{center}
\caption{\label{fig:MCexp} Transmission spectrum of a rubidium Rydberg-EIT medium as observed in experiments (a,c) and obtained from Monte Carlo calulations (b,d). (a,b) Spectra for a peak density of $\rho=1.2\times10^{10}$cm$^{-3}$, $\Omega= 4.8$ MHz and $\Omega_{\rm p}=0.3$MHz (circles, solid line), $1.0$ MHz (squares, dashed line) and $2.0$ MHz (triangles, dotted line). (c,d) Spectra for a smaller peak density of $\rho=3.5\times10^{9}{\rm cm}^{-3}$,$\Omega= 3.8$ MHz and $\Omega_{\rm p}=0.08$MHz (circles, solid line), $1.1$ MHz (squares, dashed line) and $2.0$ MHz (triangles, dotted line). Figure adapted from \cite{Sevincli2011a}.}
\end{figure}

The close connection of the Rydberg blockade and the emerging optical nonlinearities can be further revealed by considering the fraction
\begin{equation}
f_{\rm bl}=\frac{\rho_{r}^{(0)}}{\rho_{r}}-1
\end{equation}
of blockaded Rydberg excitations, where $\rho_{r}$ and $\rho_{r}^{(0)}$ are the steady state densities of excited Rydberg atoms with and without interactions, respectively. Plotting the scaled absorption coefficient ${\rm Im}(\chi)/{\rm Im}(\chi_{2l})$ against the blockaded excitation fraction reveals a universal behaviour that follows the simple relation
\begin{equation}\label{eq:scaling}
\frac{{\rm Im}(\chi)}{{\rm Im}(\chi_{2l})}=\frac{f_{\rm bl}}{1+f_{\rm bl}}\;,
\end{equation}
showing that the onset of saturation of the optical response indeed coincides with the Rydberg blockade that becomes significant when $f_{\rm bl}$ starts to exceed unity. More recently \cite{Garttner2014}, this relation has been proven to indeed hold exactly within the classical rate-equation limit \cite{Ates2011}.

Eq.(\ref{eq:scaling}) moreover reveals directly the nonlinear character of the optical susceptibility. From the form of the dark state Eq.(\ref{eq:darkstate}) we see that the Rydberg excitation probability, and therefore the blockaded fraction, scales as $f_{\rm bl}\sim|\Omega_{\rm p}|^2/\Omega^2$ at low probe intensities. Hence, a third-order nonlinearity $\chi \Omega_{\rm p}\sim |\Omega_{\rm p}|^2\Omega_{\rm p}$ emerges at small probe intensities. Owing to the enormous strength and range of typical Rydberg-Rydberg atom interactions, the excitation blockade sets in at very low light intensities, and thereby induces optical nonlinearities that are substantially stronger than previously observed in any other medium. 

Despite the simplicity of the above picture, the underlying Monte-Carlo simulations show remarkable agreement with transmission measurements \cite{Pritchard2010,Sevincli2011a} in cold rubidium Rydberg gases (FIG.\ref{fig:MCexp}), while extensions of the scaling relation Eq.(\ref{eq:scaling}) permit a quantitative interpretation of observed dispersive nonlinearities of Rydberg gases in an optical cavity \cite{Parigi2012}. The Monte Carlo simulations of the atomic gas can also be coupled to the classical evolution equation of the probe field to study nonlinear propagation effects \cite{Garttner2013}. In the following we will discuss an alternative approach to this problem that additionally reveals the highly nonlocal character of the resulting nonlinear phenomena, and serves as an insightful precursor of the discussion of quantum effects in Section \ref{sec: Quantum nonlinear optics}. 

\subsection{Nonlinear light propagation}
\label{subsec:Nonlinear light propagation}
Extending the picture developed above, we will now discuss an effective evolution equation for the coherent photon field to capture its nonlinear propagation through a Rydberg-EIT medium. Assuming that the interactions do not change the classical nature of the coherent input field (corresponding to $OD_{\rm b}\ll1$) we can also discard correlations between photons and atomic excitations and describe the light according to its amplitude defined by the expectation value $\mathcal{E}({\bf r})=\langle\hat{\mathcal{E}}({\bf r})\rangle$. Under these conditions it is possible to derive an effective propagation equation for the light field from the Heisenberg equations (\ref{eq:HB1})-(\ref{eq:HB3}). For simplicity, we will further assume that the control field Rabi frequency obeys $\Omega<\sqrt{\gamma^2+\Delta^2}$, which permits to adiabatically eliminate the polarisation dynamics in Eq.(\ref{eq:HB2}). Taking expectation values in the remaining two equations, the dynamics of the photon and Rydberg spin wave field reduce to
\begin{align}
\label{eq:clHB1}
\partial_t {\mathcal{E}}({\bf r}) & =  \left(ic\frac{\nabla_{\perp}}{2k}-c \partial_z\right){\mathcal{E}}({\bf r})-\frac{g^2\rho}{\Gamma}{\mathcal{E}}({\bf r}) -\frac{\Omega g\sqrt{\rho}}{\Gamma} \langle\hat S({\bf r})\rangle,  \\
\label{eq:clHB2}
\partial_t \langle\hat S({\bf r})\rangle & =-\frac{\Omega g\sqrt{\rho}}{\Gamma} {\mathcal{E}}({\bf r})-\frac{\Omega^2}{\Gamma}\langle\hat S({\bf r})\rangle+i \int {\rm d}{\bf r}^{\prime} V(|{\bf r}-{\bf r}^\prime|) \langle \hat S^\dagger({\bf r}^\prime) \hat S({\bf r}^\prime) \hat S({\bf r})\rangle,
\end{align}
where $\Gamma=\gamma-i\Delta$. While eqs.(\ref{eq:HB1})-(\ref{eq:HB3}) neglect transverse beam diffraction, Eq.(\ref{eq:clHB1},\ref{eq:clHB2}) account for such effects through the transverse kinetic energy term with an effective photon mass $k/c$, where $k$ is the wave number of the probe photons. For strong interactions, the photon propagation is determined by correlations between Rydberg spin wave excitations, whose dynamics can be readily obtained from the Heisenberg equations (\ref{eq:HB1})-(\ref{eq:HB3}). The dynamics of the simplest nontrivial term is given by
\begin{align}
\label{eq:SS}
\partial_t \langle\hat S({\bf r}) \hat S({\bf r}^\prime)\rangle  = & -\frac{\Omega g\sqrt{\rho}}{\Gamma}   \left({\mathcal{E}}({\bf r})\langle \hat S({\bf r}^\prime)\rangle +\langle \hat  S({\bf r})\rangle{\mathcal{E}}({\bf r}^\prime) \right)  -2\frac{\Omega^2}{\Gamma}\langle \hat S({\bf r})S({\bf r}^\prime)\rangle\nonumber\\ 
&+i V(|{\bf r}-{\bf r}^{\prime}|) \langle\hat  S({\bf r})\hat S({\bf r}^\prime)\rangle \nonumber\\
&+i\int{\rm d}{\bf r}^{\prime\prime}\left[V(|{\bf r}-{\bf r}^{\prime\prime}|)+V(|{\bf r}^\prime-{\bf r}^{\prime\prime}|)\right]\langle \hat S^\dagger({\bf r}^{\prime\prime}) \hat S({\bf r}^{\prime\prime}) \hat S({\bf r})\hat S({\bf r}^\prime) \rangle.
\end{align}
Expectedly, the propagation of such two-body correlators requires knowledge about three-body correlations as given by the last term in Eq.(\ref{eq:SS}). Eventually, this generates an infinite hierarchy of equations for an exact description of the quantum many-body dynamics and typically requires a truncation through suitable closure relations \cite{Bonitz2016}. In the present situation a simple approach derives from the fact that the Rydberg atom density $\langle\hat S^\dagger \hat S\rangle=g^2\rho/\Omega^2 |\mathcal{E}|^2$ in the absence of interactions provides an upper bound on the actual density of Rydberg atoms since the interaction blockade \cite{Lukin2001} will suppress the excitation of Rydberg states. The number $z_{\rm b}^3\langle\hat S^\dagger \hat S\rangle$ of Rydberg excitations within the interaction volume is, therefore, much less than unity in the limit of low light intensities, which is of primary concern here. Consequently, three-body interaction terms are greatly suppressed and can be neglected to leading order. As shown in \cite{Sevincli2011}, this enables an exact determination of all binary spin wave correlators in the steady state and in particular yields 
\begin{equation}\label{eq:SSS}
\langle\hat  S^\dagger({\bf r}^\prime)\hat S({\bf r}^\prime) \hat  S({\bf r})\rangle=
-\frac{g^3\rho^{3/2}}{\Omega^3}\frac{2\Omega^2/\Gamma}{2\Omega^2/\Gamma -i V(|{\bf r}-{\bf r}^\prime|)}  |{\mathcal{E}}({\bf r}^\prime) |^2 {\mathcal{E}}({\bf r}).
\end{equation}
This simple expression nicely reflects the Rydberg blockade effect, i.e. a strong suppression of two Rydberg excitations at a distance $z_{\rm b}$ for which their mutual interaction $V(z_{\rm b})$ exceeds their combined EIT line width $2\Omega^2/|\Gamma|$. 

\begin{figure}[!t]
\begin{center}
\includegraphics[width=0.95\textwidth]{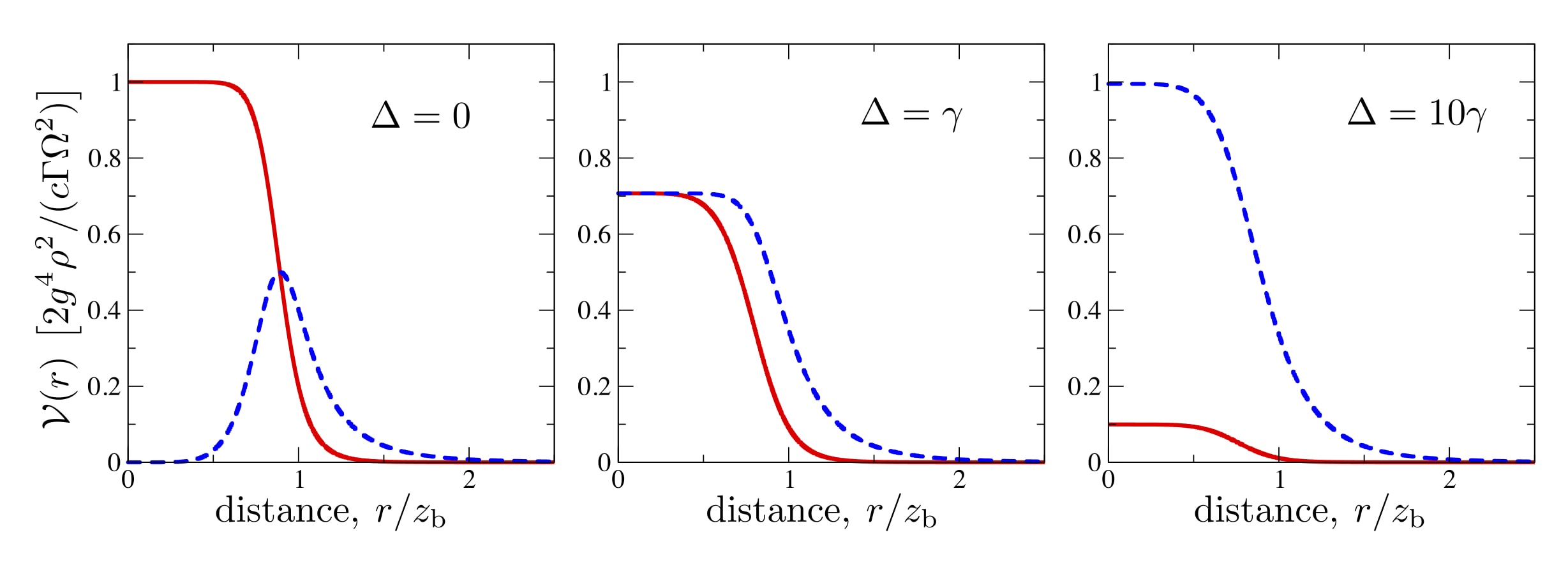}
\end{center}
\caption{\label{fig:effpot} Real (dashed curves) and imaginary (solid curves) part of the nonlinear kernel given in Eq.(\ref{eq:effpot}). Tuning the probe light onto single-photon resonance ($\Delta=0$) results in strong nonlinear absorption, while dispersive photon interactions dominate for far detuned ($|\Delta|\gg\gamma$) optical coupling.}
\end{figure}

The consequences of this blockade effect for the photon propagation are revealed by using Eq.(\ref{eq:SSS}) to obtain the steady state spin wave amplitude from Eq.(\ref{eq:clHB2}) and eventually a closed equation for the photon field
\begin{equation}
\label{eq:nonlinprop}
i\partial_z{\mathcal{E}}({\bf r})  = -\frac{\nabla_{\perp}^2}{2k}{\mathcal{E}}({\bf r}) -  \int {\rm d}{\bf r}^{\prime} \mathcal{V}(|{\bf r}-{\bf r}^\prime|)  |{\mathcal{E}}({\bf r}^\prime) |^2 {\mathcal{E}}({\bf r}),
\end{equation}
with the effective interaction potential
\begin{equation}
\label{eq:effpot}
\mathcal{V}(r)=\frac{g^4\rho^2}{c\Gamma\Omega^2} \frac{2V(|{\bf r}-{\bf r}^\prime|)}{2\Omega^2/\Gamma -i V(|{\bf r}-{\bf r}^\prime|)}.
\end{equation}
The effective potential is a complex quantity and causes nonlinear refraction as well as absorption. At large distances, it follows that $\mathcal{V}\approx\frac{g^4\rho^2}{c\Omega^4}V$, which can be interpreted to emerge from the interacting spin wave component $\hat S=-\frac{g\sqrt{\rho}}{\Omega}\hat{\mathcal{E}}$ of the underlying dark state polaritons which form, and is consistent with the expectation from a perturbative quantum treatment in the limit $V\ll\Omega^2/|\Gamma|$ \cite{Friedler2005}. With decreasing distance, however, the Rydberg blockade breaks up the polaritons and causes the photon interaction to saturate to a finite value, $\mathcal{V}(0)=i\frac{2g^4\rho^2}{c\Gamma\Omega^2}$, related to the two-level response of the medium. This behaviour is displayed in FIG.\ref{fig:effpot}, which also illustrates the tunability of the dissipative and dispersive contribution to the effective interaction via the single-photon detuning $\Delta$.

Nonlocal optical nonlinearities have also been found for a range of other materials, where the finite spatial extent of the nonlinear kernel, $\mathcal{V}$, typically arises from certain types of transport processes \cite{Ultanir2003,Dabby1968,Derrien2000,Suter1993,Skupin2007}. While the resulting kernels typically diverge at small distances, the Rydberg blockade mechanism leads to a characteristic soft core shape of the effective photon interactions with unprecedented strength that exceeds previous records by several orders of magnitude. As will be discussed below, this facilitates distinct nonlinear light propagation effects and makes them observable at extremely low light intensities.

\subsubsection{Dissipative effects}
\label{subsubsec:Dissipative limit}
\begin{figure}[!t]
\begin{center}
\includegraphics[width=\textwidth]{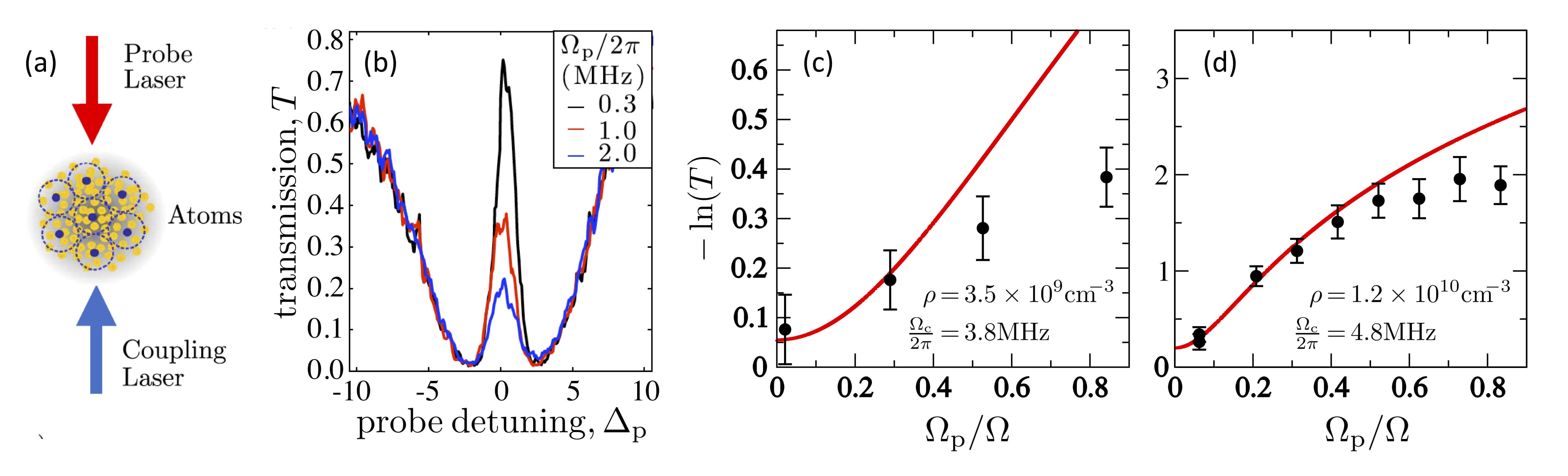}
\end{center}
\caption{\label{fig:nonlinabs} The transmission spectrum of a cold rubidium Rydberg gas (a) is shown in panel (b) for $60S_{1/2}$ Rydberg states as a function of the probe detuning at an atomic density of $\rho=1.2\times 10^{10}{\rm cm}^{-3}$ for different probe Rabi frequencies indicated in the figure. (c) and (d) The corresponding resonant transmission as a function of the probe-field amplitude, comparing experimental measurements (dots) to theoretical results based on the nonlinear propagation equation (\ref{eq:abs}) (solid lines). The dashed lines show corresponding theory results upon neglecting the drop in absorption due to attenuation and averaging over the initial transverse beam profile. Panels (a,b) and (c,d) adapted from \cite{Pritchard2010} and \cite{Sevincli2011}, respectively.}
\end{figure}

On single-photon resonance the major effect of $\mathcal{V}$ is to cause absorption. In fact, the corresponding measurement of nonlinear light absorption in a cold rubidium gas by Pritchard et al. \cite{Pritchard2010} marks the first experimental demonstration of nonlinear optical effects in interacting Rydberg gases. Here a cold gas of rubidium atoms was illuminated by a weak probe field and a resonant control field that coupled to the $60S_{1/2}$ Rydberg state (FIG.\ref{fig:nonlinabs}a). The measured probe-transmission spectrum revealed a strongly enhanced absorption near two-photon resonance, nearly extinguishing the EIT effect at the highest probe intensities shown in FIG.\ref{fig:nonlinabs}b.

For such strong absorption one can neglect transverse beam diffraction and use a local approximation of Eq.(\ref{eq:nonlinprop}) 
\begin{equation}\label{eq:abs}
\partial_z |\mathcal{E}({\bf r})|^2=-2{\rm Im}(\chi^{(3)})|\mathcal{E}({\bf r})|^4
\end{equation}
to describe the nonlinear intensity attenuation of the traversing light field. Interestingly the real and imaginary parts of the third-order susceptibility \cite{Sevincli2011}
\begin{equation}
{\rm Im}(\chi^{(3)})={\rm Re}(\chi^{(3)})=\int{\rm d}{\bf r}\mathcal{V}(r)=\frac{2}{3}\pi z_{\rm b}^3\frac{g^4\rho^2}{c\Gamma\Omega^2}
\end{equation}
coincide for van der Waals interactions, $V(r)=C_6/r^6$, and can take on giant values of $\sim0.1V^{-2}m^2$ for the experimental conditions realised in \cite{Pritchard2010}. Direct comparisons to the predictions of the simple nonlinear propagation Eq.\ref{eq:nonlinprop}, augmented to incorporate density inhomogeneities and finite laser line widths, showed remarkably good agreement with the low-intensity measurement results \cite{Ates2011} (FIG.\ref{fig:nonlinabs}c and d).

\subsubsection{Dispersive effects}
\label{subsubsec:Dispersive limit}
On resonance the strong nonlinear absorption will overshadow any dispersive effects, but can be strongly suppressed by working off single-photon resonance. For far detuned fields, $\Delta\gg\gamma$, ${\rm Re}(\chi^{(3)})\sim (\Delta/\gamma){\rm Im}(\chi^{(3)})$ such that coherent effective photon interactions can be made dominant in this limit, as also illustrated in FIG.\ref{fig:effpot}. 

The nonlinear propagation equation Eq.(\ref{eq:nonlinprop}) then becomes formally equivalent to a two-dimensional Gross-Pitaevskii equation, that, e.g., describes weakly interacting atomic quantum gases \cite{Dalfovo1999}. Here, the longitudinal propagation coordinate $z$ corresponds to an effective time with the initial state given by the input profile of the probe beam. Eq.(\ref{eq:nonlinprop}) for the present photonic system can thus be interpreted to describe a two-dimensional quantum gas evolving under finite range interactions of the type shown in FIG.\ref{fig:effpot}. While the sign of the plateau value, $\mathcal{V}(0)=-\frac{2g^4\rho^2}{c\Delta\Omega^2}$, could be conveniently controlled via the single-photon detuning, the detuning sign has to be carefully chosen in order to avoid resonances occurring at $V(r)=-2\Omega^2/\Delta$ \cite{Sevincli2011,Gorshkov2011}. Thus, one can realise defocussing nonlinearities with repulsive Rydberg atom interactions if $\Delta>0$, while focussing nonlinearities can be achieved with attractive interactions if $\Delta<0$. 

\begin{figure}[!t]
\begin{center}
\includegraphics[width=0.9\textwidth]{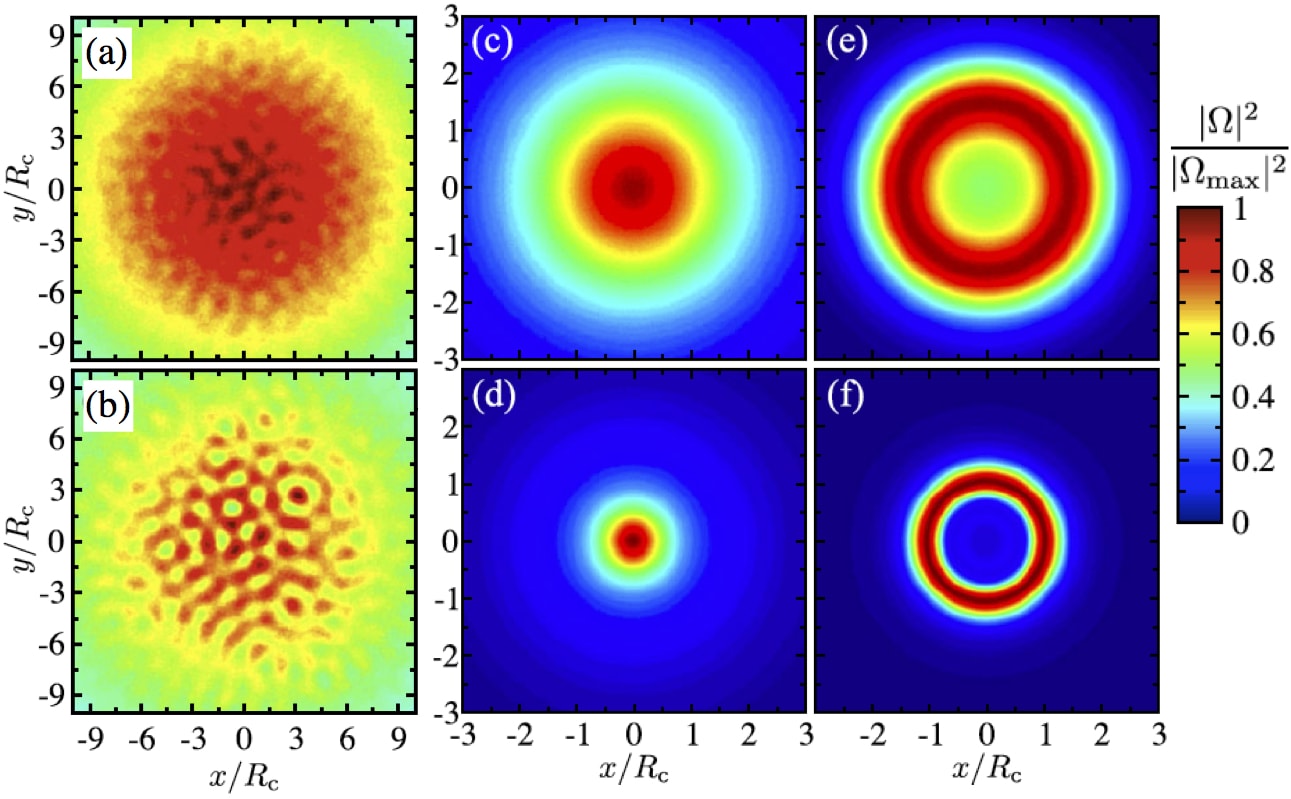}
\end{center}
\caption{\label{fig:nonlinprop} Examples of nonlinear propagation effects. (a,b) Transverse intensity profiles of the outgoing probe beam for repulsive interactions occurring in rubidium gases. The shown filaments where obtained for a super-Gaussian input beam and different densities of (a) $\rho=5.5\times10^13{\rm cm}^{-3}$ and (b) $\rho=8\times10^13{\rm cm}^{-3}$. Panels (c-f) illustrate the formation of solitons for attractive photon interactions achievable with $n^{\:1\!}S_0$ Rydberg states of strontium atoms. The results where obtained for two different densities of (c,e) $\rho=8\times10^11{\rm cm}^{-3}$ and (d,f) $\rho=1.2\times10^12{\rm cm}^{-3}$ and (c,d) a Gaussian and (e,f) super-Gaussian input beam. For the color coding each distribution has been normalized by the actual maximum squared Rabi frequency $|\Omega_{\rm max}|^2$. Figure reproduced from \cite{Sevincli2011}.}
\end{figure}

The latter case can be realised, e.g., with $n^{\:1\!}S_0$ Rydberg states of strontium atoms \cite{Mukherjee2011} for which two-photon Rydberg excitation and EIT has been demonstrated in several experiments \cite{Millen2010,Millen2011,McQuillen2013,Ye2013,DeSalvo2015,Camargo2016,DeSalvo2016,Bridge2016}. The specific soft-core form of the effective photon-photon interaction promotes the formation of solitons but, in contrast to local \cite{Weinstein1983} or some other types of nonlocal \cite{Maucher2011} nonlinearities, does not suffer from wave collapse in two dimensions. As illustrated in FIG.\ref{fig:nonlinprop}c-f, this is predicted \cite{Sevincli2011} to cause tight beam focussing or the spontaneous formation of ring structures, i.e. hollow beams, due to nonlinear light propagation.

Repulsive interactions, on the other hand, can be generated with $nS_{1/2}$ states of cold rubidium atoms, as used in the majority of Rydberg-EIT experiments. The emerging defocussing nonlinearity in this case has profound effects on the light propagation that again are distinct from other known types of  nonlinearities, whose nonlocality arises from various types of transport processes \cite{Ultanir2003,Dabby1968,Derrien2000,Suter1993,Skupin2007}. While $\mathcal{V}(r)$ is purely repulsive in real space, the corresponding Fourier space interaction $\tilde{\mathcal{V}}(k)$ is not sign definite and can turn negative over a finite interval range of wave numbers $k$. As a result, the excitation spectrum of the system can develop a roton-maxon structure that bears formal analogies to the physics of superfluid helium \cite{Landau1941,Feynman1956}. The matter-wave equivalent of this behaviour has also been predicted for dipolar \cite{Santos2003} or Rydberg-dressed \cite{Henkel2010} Bose-Einstein condensates, and can cause a roton instability that breaks continuous translational symmetry to form spatial patterns, such as those shown in FIG.\ref{fig:nonlinprop}a and b. In contrast to, e.g., dipolar systems, the soft core of the potential renders these structures stable against collapse. As the photonic interactions are isotropically repulsive, the propagation dynamics would eventually culminate in regular patterns of optical filaments in the presence of transverse beam confinement \cite{Maucher2016}.

This pattern formation process can be seen as the high density analog of photon crystallization scenarios that where recently discussed for cold Rydberg gases in optical cavities at very low photon densities or weak interactions \cite{Sommer2015}. Here, the formalism outlined in Section \ref{subsec:Nonlinear light propagation} should equivalently apply to such situations with the propagation coordinate $z$ replaced by the real time $t$. In fact, the photon interaction potential $\mathcal{V}/[g^4\rho^2/(\Omega^4c)]$ of Eq.(\ref{eq:effpot}) turns out to be identical to the effective potential derived from a quantum treatment of binary polariton scattering \cite{Bienias2014} in the limit of low energies and $\Omega\ll |\Delta|$. From this perspective, the above semiclassical theory \cite{Sevincli2011} may therefore be a promising starting point for the incorporation of quantum effects and a systematic extension to higher densities, e.g., by considering higher order interactions as derived recently in \cite{Bai2016}.

According to Eq.(\ref{eq:SSS}), the emergence of the optical nonlinearity may also be interpreted as a Rydberg-dressing of photons, formally equivalent to Rydberg-dressing of ground state atoms \cite{Henkel2010,Pupillo2010,Honer2010}, where the weak Rydberg-state admixture endows the dressed state with effective interactions. As pointed out recently \cite{Gaul2015,Helmrich2016}, driving a Rydberg gas under EIT conditions also generates effective atomic interactions, which suggests interesting dynamics involving both light- and matter-wave fields due to mutually induced nonlinearities \cite{Ostermann2016,Labeyrie2014}.

\section{Quantum nonlinear optics}
\label{sec: Quantum nonlinear optics}
Having presented a classical picture for the emergence of photonic interactions in the low-$OD_b$ limit, we will now turn focus to the regime of quantum nonlinear optics, in which the optical depth per blockade radius exceeds unity and nonlinear optical phenomena begin to affect the quantum nature of light on a few-photon level. Naturally, it now becomes necessary in this case to account for photon-photon and photon-atom correlations explicitly in order to gain a complete understanding of the underlying physics. As we shall see, strong nonlinear dissipation on single-photon resonance, to be discussed in Section \ref{sec: Dissipative quantum nonlinearity}, provides a useful mechanism for preparing and manipulating nonclassical states of light, Section \ref{sec: Preparation of nonclassical states of light}, and optical switching, Section \ref{sec: All-optical switch}. A description of dispersive effects will then be provided in Section \ref{sec: Dispersive quantum nonlinearity} for the case in which the single-photon detuning is large, including a discussion of the resulting two-photon dynamics in Section \ref{sec: Photonic bound state} and potential applications for implementing photonic phase gates in Section \ref{sec: Two-photon phase gate}.

\subsection{Dissipative quantum nonlinearity}
\label{sec: Dissipative quantum nonlinearity}
\begin{figure*}[!b]
\begin{center}
\includegraphics[width=\textwidth]{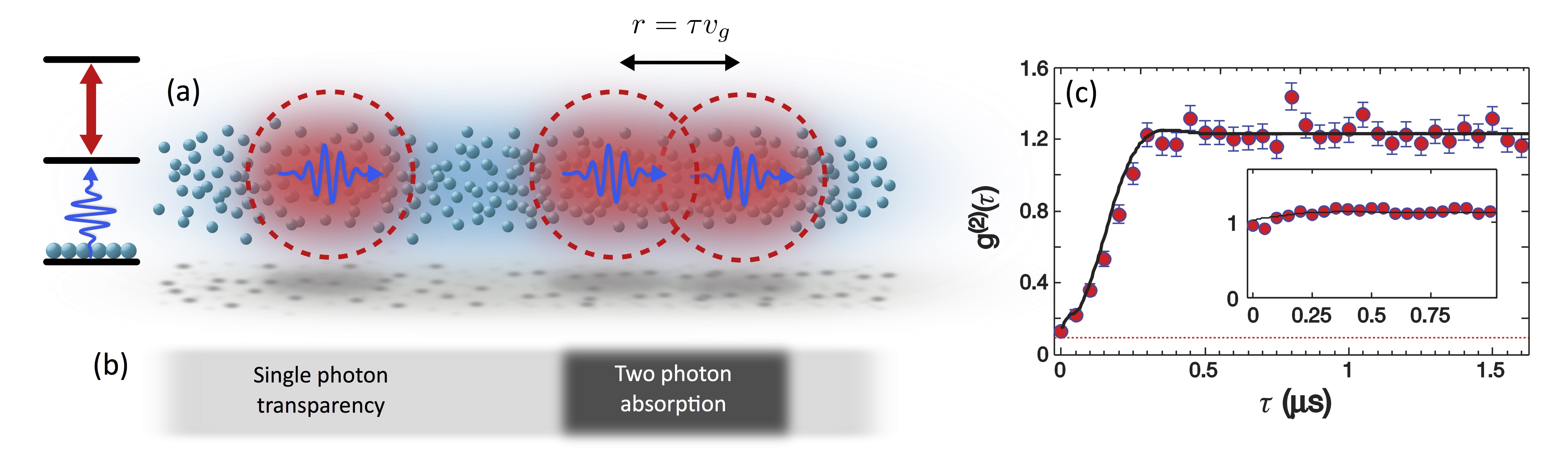}
\end{center}
\vspace{-5ex}
\caption{\label{fig: Dissipative nonlinearity} (a) The dissipative optical nonlinearity in a Rydberg EIT medium gives rise to an effective photon blockade mechanism in which well-separated photon pairs can propagate transparently as Rydberg polaritons whilst closely separated photons are absorbed, as indicated in (b). This causes photons to become anti-bunched in their spatial separation $r$ (or equivalently, in their time delay $\tau = r /v_g$). (c) Measured temporal correlation function $g^{(2)}(\tau)$ of photons transmitted through a rubidium Rydberg-EIT medium for $n=100$ (main figure) and $n=46$ (inset), highlighting the transition from classical to non-classical light with increasing $OD_b$. Panel (c) adapted from \cite{Peyronel2012}.}
\end{figure*}

Recall that on single photon resonance ($\Delta = 0$), the interatomic Rydberg interactions predominantly result in strong nonlinear probe field absorption due to the effective dissipative two-level medium established within the blockade volume surrounding a Rydberg polariton. Within a quantum picture, this dissipative nonlinearity may then be expected to realise a photon blockade mechanism, allowing single photons to propagate transparently, whilst strongly absorbing closely separated pairs of photons, FIG.\ref{fig: Dissipative nonlinearity}a and b. The initial realization of this kind of nonlinearity, reported by \cite{Peyronel2012}, marked the first demonstration of a quantum optical nonlinearity by means of Rydberg EIT.

This simple picture suggests that any two photons propagating within a distance $z<z_b$ from each other cannot simultaneously traverse the medium provided the optical depth per blockade radius satisfies $OD_b>1$. As a result, the transmitted light is expected to be highly nonclassical with an equal-time correlation function $g^{(2)}(0) \approx0$ \cite{Petrosyan2011}. We remark that the above classical theory was built on the assumption $g^{(2)}(0)=1$ \cite{Sevincli2011}, which makes its good agreement with the experiments by Pritchard et al. \cite{Pritchard2010} seem rather surprising in retrospect. As we shall discuss below, this seeming contraction can only be resolved within a quantum treatment of light propagation and traced back to the relative motion of strongly interacting photons.

Let us base the discussion on a wavefunction treatment for the simplest scenario of two interacting photons \cite{Gorshkov2011}, which is sufficient to gain a basic insight into the aforementioned anti-bunching behaviour of light. The two-photon component of the wavefunction in this case can be written in the following manner 
\begin{equation}
| \Psi (t) \rangle = \frac{1}{2} \int dz_1 \int dz_2 EE(z_1, z_2, t) \hat{\mathcal{E}}^{\dagger}(z_1, t) \hat{\mathcal{E}}^{\dagger}(z_2, t) |0\rangle.
\end{equation}
where $\hat{\mathcal{E}}^{\dagger}(z_1)$ and  $\hat{\mathcal{E}}^{\dagger}(z_2)$ respectively create probe photons at positions $z_1$ and $z_2$, whilst $EE(z_1, z_2, t)$ is the associated time-dependant amplitude of the photon pair. For very long probe pulses it suffices to consider an adiabatic scenario, i.e., the photonic steady state defined by $\partial_t EE=0$. In this case, it can be shown \cite{Peyronel2012}, that the two-photon amplitude approximately obeys a remarkably simple equation of the following form
\begin{equation}
\label{eq:Effective diffusion equation}
\partial_R EE(R, r) = - \frac{\mathcal{V}(r)}{l_{\text{abs}}} EE(R, r) + 4 l_{\text{abs}} \left[ 1 + \mathcal{V}(r) \frac{\Omega^2}{\gamma^2} \right] \partial_r ^2 EE(R, r) 
\end{equation}
where $r$ and $R$ are the relative and centre of mass coordinates of the photon pair, $l_{\text{abs}}$ is the resonant two-level absorption length and $\mathcal{V}(r)$ is an effective potential which characterises the complex optical response of the medium, defined according to
\begin{equation}
\mathcal{V}(r) = \frac{1}{1-2i(r/z_b)^6}
\end{equation}
In the same spirit as Eq.(\ref{eq:effpot}), this potential is vanishing for $r>z_b$ and saturates to unity for $r<z_b$, which leads to a distance-dependent modification to the optical response of the medium.

The effective diffusion in the relative coordinate between photons as their centre of mass moves through the medium can be traced back to the finite EIT bandwidth of the system that leads to an absorptive filtering of small length-scale features in two-photon amplitude. Ultimately, it is the competition between this diffusive behaviour and the strong two-photon absorption for $r<z_b$ which determines the extent to which an incoming coherent photon field is converted into nonclassical light. The essential figure of merit readily follows by rewriting Eq.(\ref{eq:Effective diffusion equation}) in terms of the scaled coordinates $\tilde r=r/z_b$ and $\tilde R=R/z_b$
\begin{equation}
\label{eq:diffscaled}
\partial_{\tilde R} EE = - 2OD_b\mathcal{V}(\tilde r) EE + 2OD_b^{-1} \left[ 1 + \mathcal{V}(\tilde r) \frac{\Omega^2}{\gamma^2} \right] \partial_{\tilde r} ^2 EE\;.
\end{equation}
which shows directly that diffusion dominates the small-distance absorption for $OD_b\ll1$ and thereby sustains the coherent nature of the illuminated light with $g^{(2)}(0)=1$. On the other hand, the first term becomes significant for $OD_b>1$ and leads to strong antibunching of the transmitted photons. This transition was indeed observed experimentally \cite{Peyronel2012}, showing nearly uncorrelated light for moderately excited Rydberg states (inset of FIG.\ref{fig: Dissipative nonlinearity}c) and virtually complete antibunching for high lying Rydberg states for which $OD_b>1$. We remark though that the temporal extent of the generated photon correlations can greatly exceed the simple estimate, $z_b / v_g$, corresponding to photons be anti-bunched over a blockade radius (where $v_g$ is the slow-light group velocity), and increases linearly with the inverse EIT bandwidth. This is ultimately by virtue of the EIT related absorptive filtering described above.

\subsubsection{Preparation of nonclassical states of light}
\label{sec: Preparation of nonclassical states of light}
\begin{figure}[!t]
\begin{center}
\includegraphics[width=\textwidth]{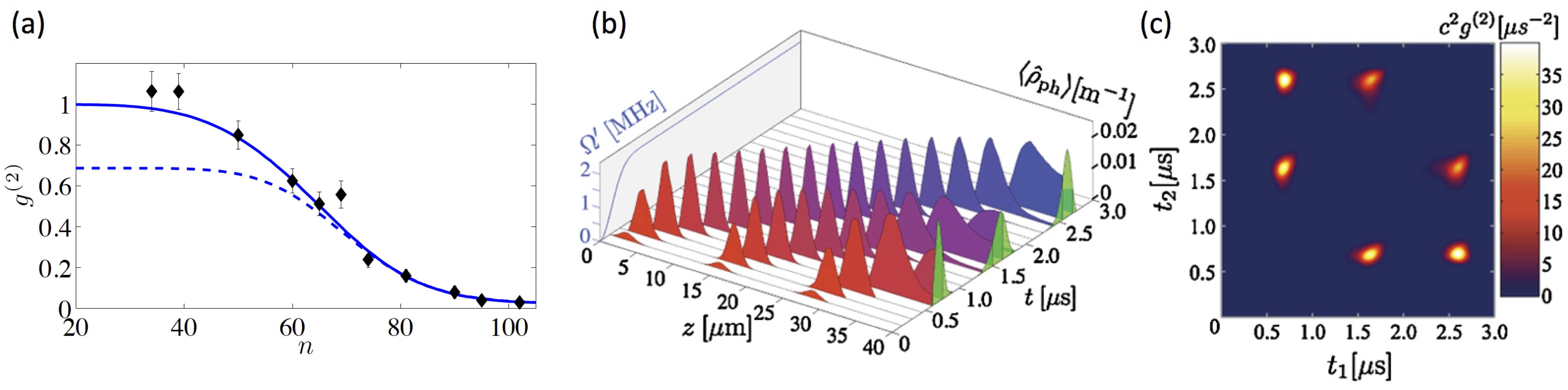}
\end{center}
\caption{\label{fig: Rydberg atoms to light} (a) Measured (dots) equal-time two-photon correlations, $g^{(2)}(0)$, of photons retrieved from stored Rydberg spin wave excitations as a function of their principle quantum number $n$ \cite{Dudin2012}. This is compared to theoretical predictions derived from the dephasing dynamics of the full $N$-body wave function (solid line) and of a restricted state space with less than three Rydberg excitations (dashed line) \cite{Bariani2012}. As $n$ increases, the blockade radius enlarges until it engulfs the entire medium, thereby admitting only a single collective excitation and, in turn, a single retrieved photon. (b) The simulated photon dynamics of light retrieved from an adiabatically prepared Rydberg atom crystal, and (c), the predicted two-photon temporal correlation function of the retrieved light \cite{Pohl2010}. Panel (a) reproduced from \cite{Bariani2012} and panels b and c reproduced from \cite{Pohl2010}.}
\end{figure}

The capacity for Rydberg based systems to generate non-classical states of light was first brought to attention in \cite{Lukin2001}, with the initial discussion of the Rydberg blockade mechanism. The idea here was to prepare an entangled spin wave state with classical light, and map the nonclassical correlations promoted by the Rydberg blockade onto photons using EIT \cite{Fleischhauer2000}. Saffmann et al. showed how this idea could be employed to use a fully blockaded ensemble as a source of single photons \cite{Saffman2002}. In 2012, Dudin et al. \cite{Dudin2012} realised this scheme, demonstrating the generation of single photon pulses with a vanishing $g^{(2)}(0)$ as shown in FIG.\ref{fig: Rydberg atoms to light}a. 

This general philosophy could also be expanded to many-body states in extended media beyond the blockade radius. For instance, adiabatic preparation schemes where predicted to yield ordered configurations of collective Rydberg-spin wave excitations \cite{Pohl2010,Schachenmayer2010,VanBijnen2011}, and recently realised experimentally \cite{Schauss2015}. Such ordered states of Rydberg excitations can be mapped onto quantum light \cite{Pohl2010} using equivalent state transfer techniques \cite{Fleischhauer2000}, resulting in a regular train of single photon pulses as shown in FIG.\ref{fig: Rydberg atoms to light}b and c.

\begin{figure}[!t]
\begin{center}
\includegraphics[width=0.95\textwidth]{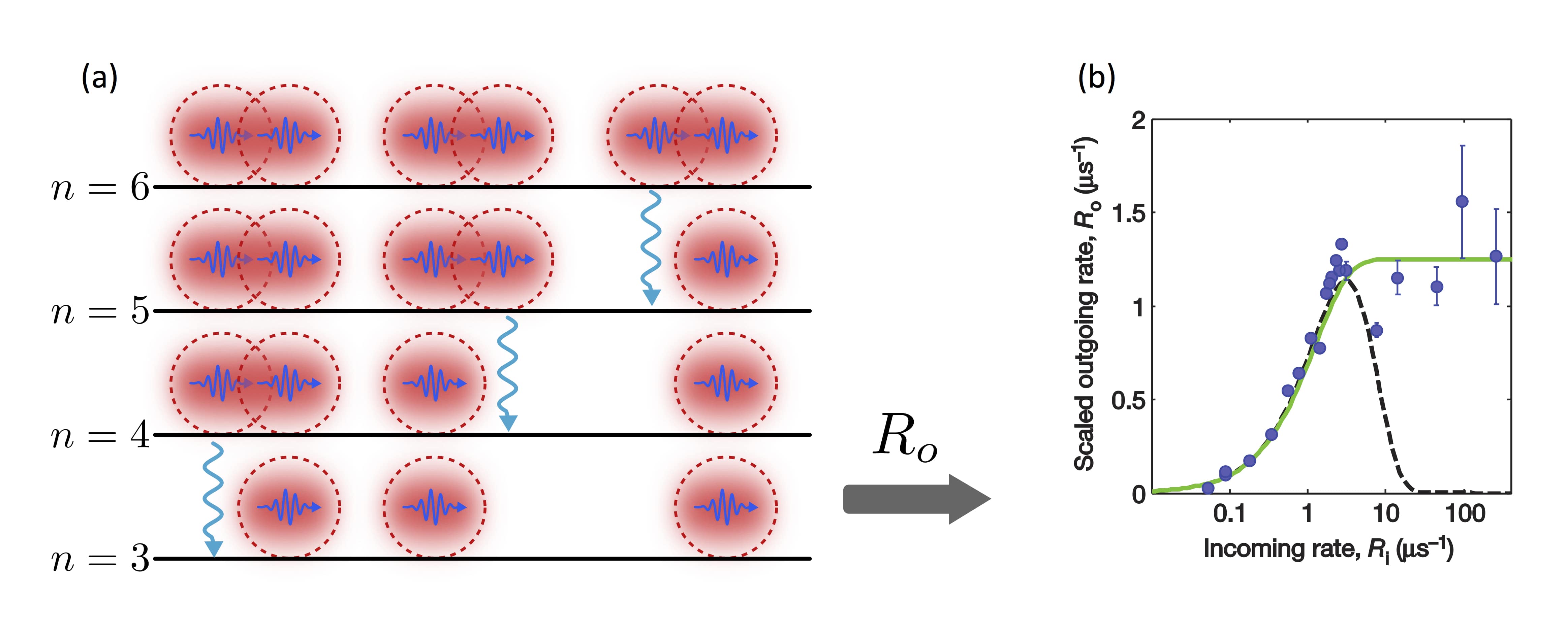}
\end{center}
\caption{\label{fig: Dissipative dynamics}(a) The process of photon scattering in a dissipative quantum nonlinear medium continually causes an incident $n$-photon state to decay towards a dilute phase in which all photons are separated by a minimum distance. This maps the incoming light onto a finite out-going photon rate of $R_{\text{o}}$. (b) The out-going photon rate is plotted as a function of the incoming rate $R_{\text{in}}$, and is shown to saturate to a value corresponding to one photon in the medium at a time, in stark contrast to the expectation from a multi-photon absorber (dashed lines). Panel (b) adapted from \cite{Peyronel2012}.}
\end{figure}

On the other hand, the availability of strong photon-photon interactions suggests a more direct and dynamical approach to generating such regular states or photon crystals from freely propagating light quanta. In the dissipative case, the formation of such exotic states would be enabled by the fact that photons are only allowed to propagate transparently every blockade radius, whilst all other photons will be absorbed. Following this picture, an incident $n$-photon pulse is expected to continually decay into $(n-1)$-photon states until reaching a dilute phase where all photons are sufficiently far away from each other such that absorption eventually comes to a halt, FIG.\ref{fig: Dissipative dynamics}a. Indeed, the experiments by Peyronel et al. \cite{Peyronel2012} demonstrated this effect by observing a saturation in the rate $R_{\text{o}}$ at which photons were transmitted through the medium when the incoming photon rate $R_{\text{in}}$ was increased. This is shown in FIG.\ref{fig: Dissipative dynamics}b and, importantly, deviates from the expected behaviour of a multi-photon absorber that would only transmit the vacuum and single-photon component of the incoming light field. The average time delay between transmitted photons, $R_{\text{o}}^{-1}$, however, was substantially larger than the correlation time of the observed $g^{(2)}(\tau)$ (see FIG.\ref{fig: Dissipative nonlinearity}), which, consequently does not exhibit any features beyond short-ranged ``nearest-neighbour" correlations. While long-range ordered photon states still await experimental verification, theoretical work has made intriguing predictions about the formation of regular photon crystals by dispersive nonlinearities based on a perturbative treatment of Rydberg-Rydberg atom interactions \cite{Otterbach2013,Moos2015}. 

Yet, strong interaction effects can be illuminated on a many-body level for short photon pulses whose EIT-compressed length lies below the blockade radius \cite{Gorshkov2013}. To this end, let us consider an incident pulse of $n$ photons in a single mode with a temporal envelope $h(t)$ and express the intial $n$-photon wavefunction in the following form
\begin{equation}
|\Psi(t)\rangle = \sqrt{n!} \int \limits_{z_1>...>z_n} \left[ \prod_{i=1}^n dz_i c^{-1/2} h(t-z_i/c) \hat{\mathcal{E}}^{\dagger}(z_i)  \right] |0\rangle.
\end{equation}
which implies an ordering of photon arrival times at the medium. 
If the EIT compressed pulse length is shorter than the blockade radius, as soon as the first photon ($i=1$) enters the medium and forms a Rydberg polariton, it will  establish an absorbing medium for all $n-1$ subsequent photons. Assuming perfect EIT-conditions and that these photons are scattered into some other spatial mode $\hat{Q}(z)$, the state of the $n$-body wavefunction after the pulse has entered the medium readily follows as \cite{Gorshkov2013}
\begin{equation}
|\Psi(t)\rangle = \sqrt{n!} \int \limits_{z_1>...>z_n} \left[ \prod_{i=1}^n dz_i c^{-1/2} h(t-z_i/c) \hat{Q}^{\dagger}(z_i)  \right] |\Phi (t) \rangle.
\end{equation}
where $|\Phi(t)\rangle$ denotes the state of the Rydberg polariton established by the first photon. 
The reduced density matrix $\rho_1(z,z^\prime)$ of the first and only transmitted photon can then be obtained by tracing $|\Psi(t)\rangle \langle \Psi(t)|$ over the $n-1$ scattered photons in $\hat{Q}(z)$ to yield,
\begin{equation}
\rho_1(z, z^\prime) =n  \rho^{(0)}_1(z, z^\prime) \left[ \int\limits_{\text{min}(z,z^\prime)/c}^{\infty} {\rm d}xh^2(t) \right]^{n-1},
\end{equation}
where $\rho^{(0)}_1$ denotes the initial density matrix of an incident photon and the remaining terms encapsulate all effects of many-body photon scattering induced by the Rydberg blockade. 

\begin{figure}[!t]
\begin{center}
\includegraphics[width=0.9\textwidth]{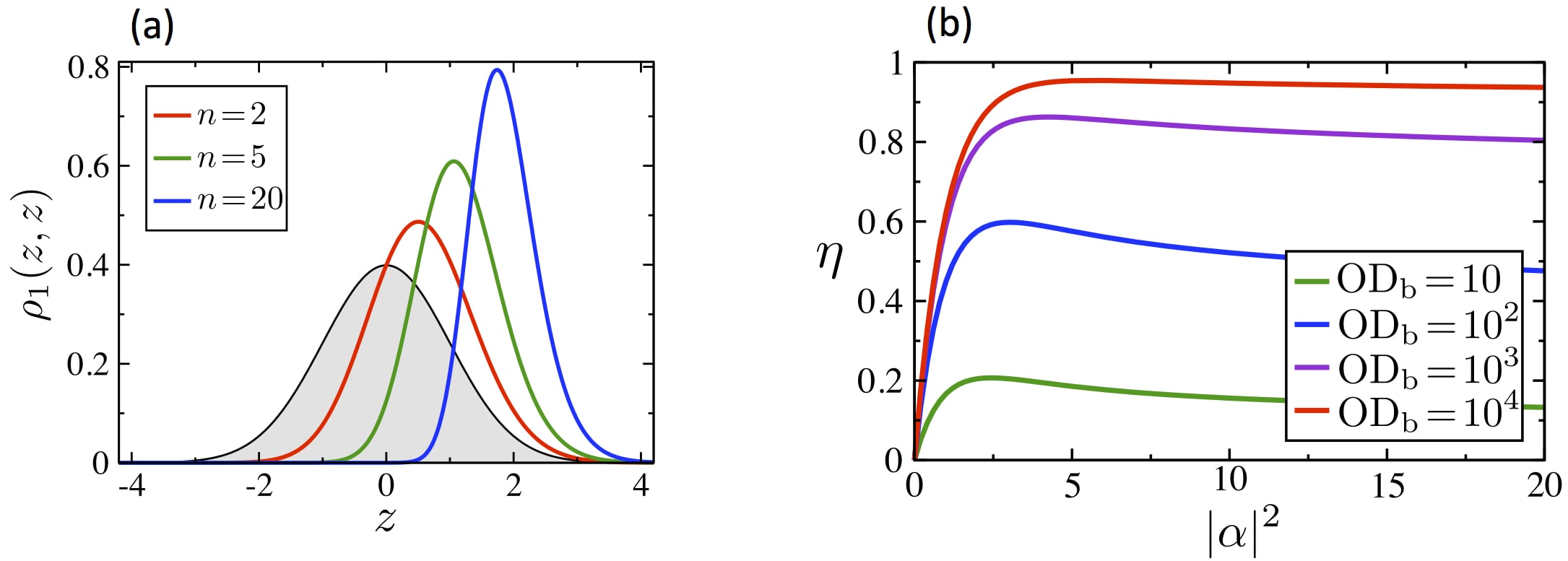}
\end{center}
\caption{\label{fig: Single photon source}(a) Assuming an incident $n$-photon Fock state, the intensity profile $\rho_1(z,z)$ of the produced single photon is shown for various indicated values of $n$, where the grey shaded curve indicates the intensity profile of the incident field. (b) The efficiency of single-photon generation is shown as a function of the mean number of photons $|\alpha|^2$ in the incident coherent-state  light field for various indicated values of $OD_b$. Panels (a) and (b) adapted from \cite{Gorshkov2013}.}
\end{figure}

As illustrated in FIG.\ref{fig: Single photon source}a, the latter causes a narrowing and significant pulse advance that grows with the number of incident photons. This behaviour results from the fact that the nonlinear light absorption relies on the presence of a Rydberg polariton, such that the first scattering event then projects the leading photon inside the medium. Since the average time for the first of such scatterings to occur decreases with the number of incident photons, the extent of this dissipative photon advance increases with $n$.
Since the trace ${\rm Tr}(\rho_1)=\int{\rm d}z\rho_1(z,z)$ is unity, the blockaded ensemble indeed acts as a perfect single-photon source under ideal EIT-conditions. However, it emerges from the medium impure. Again, this is because the scattering of all later photons is conditioned on the first photon already being in the medium, so these scattered photons ultimately carry information about the produced single photon. Such a degradation of coherence is inherent to the dissipative nature of the nonlinearity and equivalently affects proposed applications for single-photon subtraction \cite{Honer2011} and optical switching \cite{Gorshkov2011,Baur2014,Tiarks2014,Gorniaczyk2014,Gorniaczyk2016}, to be discussed below. Remarkably, in the present case the purity of the outgoing photon is $\text{Tr}[\rho_1^2(z, z^\prime)] = n/(2n - 1)$, independent on the mode function $h(t)$ and, crucially, does not vanish in the high-intensity limit $n \to \infty$.

While the above discussion provides an appealingly simple understanding of a nontrivial many-body problem, it yet draws a rather idealised picture, since in reality the finite EIT bandwidth of the system considerably influences the efficiency of such a single photon source under typical operation conditions. Although a single photon is indeed produced inside the medium with near unit fidelity, the ensuing EIT losses are what limits its ability to finally exit the medium. Indeed, the condition of a short photon pulse as well as the projection-induced pulse narrowing both require a large $OD_b$ in order to minimise such losses, as illustrated in FIG.\ref{fig: Single photon source}b.

\subsubsection{All-optical switching}
\label{sec: All-optical switch}

An optical switch is a medium through which the transmission of one (target) field can be regulated or blocked by the application of a second (gate) field. The fundamental limit of such a device is a single photon switch, in which only a single incoming gate photon is sufficient to extinguish the target field transmission. Significant efforts have been directed towards reaching this regime, as such single-photon switching capabilities lie at the heart of many quantum communication and computation protocols, and could be exploited for non-demolition measurements of single light quanta \cite{chang2007}. 

Early implementations of optical switches were operational with up to a few hundred gate photons, and exploited nonlinearities arising from the strong coupling of light to either single molecules  \cite{Hwang2009}, quantum dots \cite{Englund2012, Bose2012, Volz2012, Loo2012} or atomic ensembles \cite{Bajcsy2009}. The regime of single-photon switching was first achieved in a cavity QED setting \cite{Chen2013}, where a single photon stored in a cavity embedded atomic ensemble was made to diminish the transmission of several hundred target photons. 

\begin{figure}[!h]
\begin{center}
\includegraphics[width=0.7\textwidth]{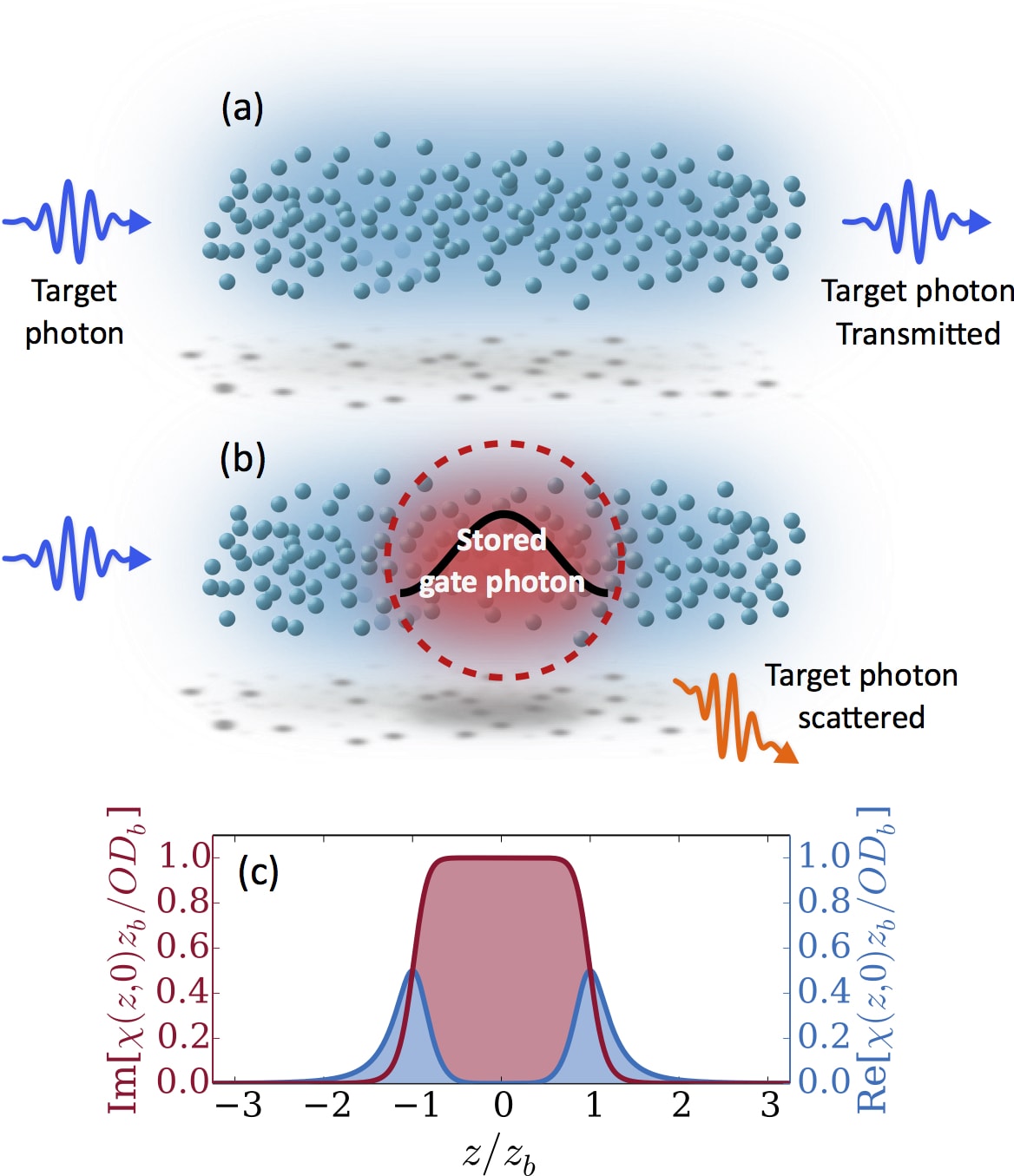}
\end{center}
\caption{\label{fig: Single photon switch} Schematics of an optical switching scheme based on collective Rydberg excitations. (a) Switch off: in the absence of a stored gate photon, the medium is transparent to the target photon. (b) Switch on: the presence of a stored gate photon causes scattering of incoming target photons, thus blocking their transmission through the medium. (c) The real (blue) and imaginary (red) parts of the medium's optical susceptibility are shown as a function of the distance from a stored Rydberg impurity atom at $z=0$, characterising the dispersive and absorptive properties of the medium respectively.}
\end{figure}

The dissipative optical nonlinearity available in the Rydberg EIT setting under current consideration provides an alternative route towards single photon switching \cite{Gorshkov2011} and has recently been implemented in a number of experiments \cite{Tiarks2014, Gorniaczyk2014,Baur2014}. The basic idea of the underlying scheme is illustrated in FIG.\ref{fig: Single photon switch}. First, a single gate photon is stored \cite{Fleischhauer2000} as a collective spin wave excitation in some Rydberg state $|r^{\prime}\rangle$ of the ensemble atoms. Subsequently, the target photons are then made to traverse the medium under EIT conditions with a different Rydberg state $|r\rangle$. Whilst the target field would propagate unimpeded in the absence of the stored gate photon, the state $|r\rangle$ experiences a level shift due to its van der Waals interaction with the Rydberg state $|r^{\prime}\rangle$ of the gate spin wave. The resulting  breaking of EIT-conditions thereby switches the medium from highly transparent to strongly absorptive. Such a spinwave conditioned absorption can also permit optical imaging of Rydberg excitations \cite{Gunter2012} as demonstrated in recent experiments \cite{Gunter2013}.

Since the interaction-induced absorption is independent of the precise state of the stored gate spin wave, it can as well be treated as a classical Rydberg impurity located at a position $z^{\prime}=0$ at the centre the medium. With this simplification, the corresponding Heisenberg equations of motion characterising the propagation of the target photons through the medium can be recast more succinctly in frequency space as a single propagation equation for the target field operator 
\begin{equation}
\label{propagation equation}
\partial_z \tilde{\mathcal{E}}(z, \omega) = i \chi \left(z, \omega \right) \tilde{\mathcal{E}}(z, \omega),
\end{equation}
where $\tilde{\mathcal{E}}(z, \omega)$ denotes the Fourier transform of $\hat{\mathcal{E}}(z, t)$ and $\chi \left(z, \omega \right)$ is the susceptibility of the medium \cite{Gorshkov2011}
\begin{equation}
\label{eq: Spin wave susceptibility}
\chi(z, \omega) = \frac{g^2\rho}{c} \frac{  \omega - V(z) }{\Omega^2 -\left(\omega + i\gamma \right) \left( \omega - V(z) \right)}.
\end{equation}
In terms of the rescaled coordinate $\tilde{z}=z/z_b$, the static contribution assumes the simple form 
\begin{equation}
\label{Resonant stored spin wave chi}
\begin{split}
\chi(\tilde{z}, 0) \approx & ~ OD_b \left[ i  \frac{1}{1 + \tilde{z}^{12}} -  \frac{\tilde{z}^6}{1 + \tilde{z}^{12}} \right],
\end{split}
\end{equation}
which is shown in FIG.\ref{fig: Single photon switch}c and exhibits the same characteristics as the effective potential derived in Section \ref{subsec:Nonlinear light propagation}. For target photon pulses with a narrow bandwidth well below the EIT linewidth, one can use Eq.(\ref{Resonant stored spin wave chi}) in Eq.(\ref{propagation equation}) to obtain a simple expression for the transmitted number of target photons \cite{Gorshkov2011}
\begin{equation}
\label{eq: Transistor solution}
\bar{N}_{\rm out} \approx \bar{N}_{\rm in} e^{ - 2 \eta}
\end{equation}
in terms of the photon number, $\bar{N}_{\rm in}$, contained in the incident target pulse. For long media with a length $L\gg z_{\rm b}$ the suppression constant
\begin{equation}
\eta  = OD_b \int_{-\infty}^\infty \frac{{\rm d}\tilde{z}}{1+\tilde{z}^{12}}\approx2OD_b
\end{equation}
becomes independent of $L$ and is consistent with the expectation that the gate spin wave exposes an absorptive two-level medium with a total extent of $\sim2z_{\rm b}$, FIG.\ref{fig: Single photon switch}c. Once more, one finds the requirement of $OD_b\gg1$ in order to achieve efficient optical switching.

\begin{figure}[!b]
\begin{center}
\includegraphics[width=\textwidth]{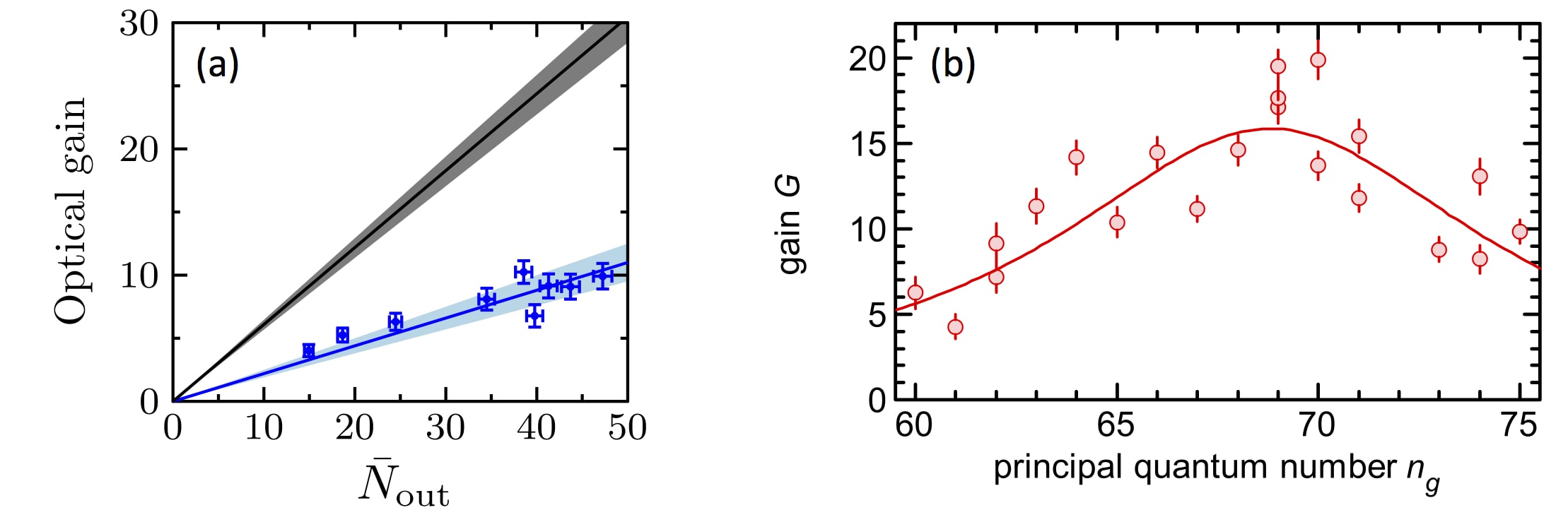}
\end{center}
\caption{\label{fig: Transistor data} (a) Optical gain of the photon switch as a function of the number of transmitted target photons, measured for a coherent gate input with an average number of $0.75$ photons (symbols) and extrapolated to the case of a single stored Rydberg impurity atom (upper line). (b) Optical gain as a function of the principal quantum number $n_g$ of the stored Rydberg spin wave. Choosing the target Rydberg state with $n_t = n_g - 2$, the gain clearly benefits from a F\"orster resonance occurring at $n_g = 69$. Panels (a) and (b) adapted from \cite{Gorniaczyk2014} and \cite{Tiarks2014} respectively.}
\end{figure}

While the optical gain under perfect EIT-conditions increases linearly with the incoming number of target photons, a practical limitation arises from the fact that target photons additionally experience self-interactions amongst themselves. These self-interactions eventually saturate the transmitted photon number $\bar{N}_{\rm out}^{(0)}$ in the absence of the gate photon and thereby limit the optical gain $G=\bar{N}_{\rm out}^{(0)}-\bar{N}_{\rm out}$. However, as demonstrated in \cite{Gorniaczyk2014, Tiarks2014} (FIG.\ref{fig: Transistor data}a and b), this limitation can be alleviated somewhat with a judicious choice of the two Rydberg states, $|r\rangle$ and $|r^\prime\rangle$. In particular, by exploiting a F\"orster resonance between these states, it is possible to minimise the self interactions between target photons yet maintain a strong interaction between atoms in $|r\rangle$ and $|r^\prime\rangle$ \cite{Saffman2009}. This ultimately enhances the achievable optical gain, as clearly illustrated in FIG.\ref{fig: Transistor data}b. 

Finally, it must be emphasised that the dissipative nature of the underlying switching mechanism fundamentally prevents the realization of a quantum transistor and restricts applications to the domain of classical switching. Analogously to the discussion of the previous section, the scattering of the target photons, into some other optical mode, serves as a measurement \cite{Gorshkov2013} of the presence of the gate spin wave, which is of course a detriment to any applications reliant on quantum coherence, such as quantum repeater protocols \cite{Briegel1998} or the generation of entangled states \cite{Gheri1997}. 

On the other hand, coherent switching capabilities are inherent to cavity-QED settings, where target photons are back-reflected from the cavity mirrors \cite{Reiserer2013, Chen2013}, or nanophotonic devices where scattered photons are returned to the well-defined fibre modes \cite{Tiecke2014, OShea2013,Shomroni2014}.  In order for future Rydberg-based transistors to be capable of operating coherent photon switching functionalities, hybrid strategies involving such strong mode confinement may be required, or perhaps new schemes altogether.

\subsection{Dispersive quantum nonlinearity}
\label{sec: Dispersive quantum nonlinearity}
\begin{figure*}[!t]
\begin{center}
\includegraphics[width=0.95\textwidth]{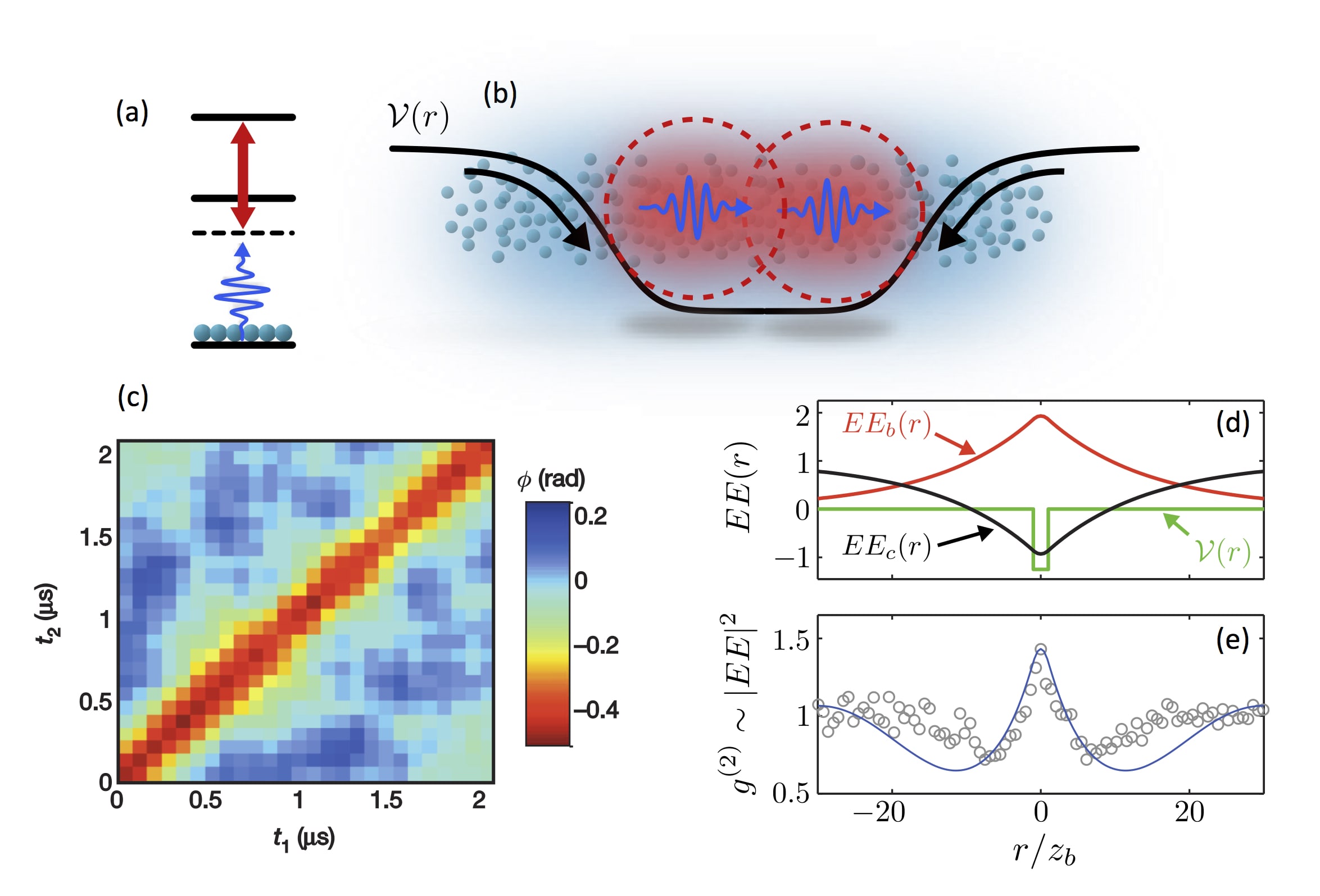}
\end{center}
\caption{\label{fig: Dispersive nonlinearity}(a) A dispersive optical nonlinearity is realised when the probe frequency is far detuned from the intermediate-state transition. (b) In this case, the refractive index of the medium is modified when two photons approach one another, which gives rise an effective photon interaction $\mathcal{V}(z)$. (c) This results in a conditional phase shift with strong spatial correlations, carrying over to a strong dependence on the photon detection times $t_1$ and $t_2$ in the experiment \cite{Firstenberg2013}. (d) The attractive potential $\mathcal{V}(z)$ supports the formation of a two-photon bound state. The initial two-photon wave-function $EE(R=0, r)=EE_0$ is a superposition of the bound state mode $EE_b$ and the continuum of scattering states $EE_c$, which acquire a relative phase as the two photon propagate through the medium. (e) Evidence for the bound state population can, hence, be observed as photon bunching in the second-order correlation function $g^{(2)}$ of the transmitted light. The line shows the theoretically predicted behaviour from $EE(R, r)$. Panels (c), (d) and (e) adapted from \cite{Firstenberg2013}.}
\end{figure*}

By detuning the frequency of photons from the intermediate two-level resonance, absorption can again be sufficiently suppressed to realise a largely dispersive nonlinearity (FIG.\ref{fig: Dispersive nonlinearity}a and b). The first experimental observation of the associated coherent interaction between propagating photons was reported in \cite{Firstenberg2013} (FIG.\ref{fig: Dispersive nonlinearity}). Besides imparting a sizeable phase on two adjacent photons (FIG.\ref{fig: Dispersive nonlinearity}c) the strong interactions give rise to interesting propagation dynamics. In particular, the spatially dependant nature of the refractive index in this case (FIG.\ref{fig: Dispersive nonlinearity}b) gives rise to an effective coherent optical potential, allowing photons to essentially acquire the characteristics of massive particles. As we shall see, this can foster the formation of exotic quantum states of light such as photonic bound states. 

\subsubsection{Photonic molecules}
\label{sec: Photonic bound state}
In close equivalence to the discussion of Section \ref{sec: Dissipative quantum nonlinearity}, the quantum dynamics of two dispersively interacting photons can be understood in terms of an effective equation for the steady state two-photon amplitude $EE(\tilde{R},\tilde{r})$. In the far detuned limit, where dissipative effects can be neglected, this takes the form of an effective Schr\"odinger equation \cite{Firstenberg2013}
\begin{equation}
\label{eq:Two photon Schrodinger equation}
i \partial_{\tilde{R}} EE(\tilde{R},\tilde{r}) = 2 \overline{OD_b}^{-1}\left[  1 - \mathcal{V}(\tilde{r}) \frac{\Omega^2}{\Delta^2} \right] \partial_{\tilde{r}}^2 EE(\tilde{R},\tilde{r}) + 2 \overline{OD}_b \mathcal{V}(\tilde{r}) EE(\tilde{R},\tilde{r})
\end{equation}
with 
\begin{equation}
\mathcal{V}(\tilde{r}) = \frac{1}{ 1 + 2\tilde{r}^6 },
\end{equation}
and where $\overline{OD}_b=(\gamma/\Delta)OD_b$ denotes the blockaded optical depth of the now off-resonant two-level medium exposed by the Rydberg blockade. Contrary to the diffusive behaviour discussed in Section \ref{sec: Dissipative quantum nonlinearity}, the first term here is equivalent to an effective kinetic energy with a distance-dependent mass $m(r) \propto - 1 / \Delta$. In principle, this allows one to to control the sign of the mass via the single-photon detuning, alluding to the possibility of engineering either attractive or repulsive effective interactions. However, the simultaneous requirement of $C_6\Delta\geq 0$, discussed in Section \ref{subsubsec:Dispersive limit}, means that the effective interaction potential $2\overline{OD}_b \mathcal{V}(\tilde{r})$ experienced by photons is such that the emergent photonic interaction is always attractive. However, a scattering analysis of two dark state polaritons \cite{Bienias2014} found that the effective interaction can be turned repulsive in the strong driving limit, $\Omega > |\Delta|$. As shown in \cite{Gaul2015} this effect stems from an interaction induced resonance with an entangled two-atom state ($|pp\rangle+|ep\rangle+|pe\rangle$) that occurs at $\Omega = |\Delta|$ and reverses the sign of the optical response at small distances.

Most remarkably, the effective Schr\"odinger equation in Eq.(\ref{eq:Two photon Schrodinger equation}) features bound state solutions. Firstenberg et al.  \cite{Firstenberg2013} succeeded to observe such exotic photon-molecules through two-photon correlation measurements that revealed strong photon bunching, as shown in FIG.\ref{fig: Dispersive nonlinearity}. Since the experiment predominantly involved only two interacting photons, bound states could not form through more common collision channels. Instead they are directly populated upon pulse entry, where the incoming two-photon state is projected onto a superposition of the bound state and a continuum of scattering states. Since these two types of solutions possess separated energies, the ensuing propagation through the medium causes a beating of the bound and continuum states, eventually resulting in a relatively enhanced contribution from the bound state for closely separated photons, and bunching behaviour in the correlation function $g^{(2)}(\tau)\sim|EE(R=L,v_{\rm g}\tau)|^2$. The limited values of $\overline{OD}_b$, realised in \cite{Firstenberg2013}, were only large enough to support a single shallow bound state with a spatial extent comparable to the size of the medium.

A more detailed analysis \cite{Bienias2014}, however, showed that this system in principle features a complex spectrum of bound state solutions. In particular, working at $C_6\Delta<0$ permits to a realize Coulomb-type photon interaction \cite{Maghrebi2015} that in turn can give rise to a hydrogen-like Rydberg series for the photons. Whilst the formation of photonic bound states has long been investigated for locally nonlinear systems \cite{Chiao1991, Deutsch1992, Deutsch1993, Drummond1997, Kheruntsyan1998, Shen2007}, work on Rydberg-EIT has enabled the first experimental observation of such exotic states of light. The produced bi-photon states can be seen as a minimal precursor to optical quantum solitons \cite{Lai1989,Lai1989a,Haus1990,Drummond1993, Kheruntsyan1998}. In fact, the described Rydberg-EIT setting could enable a systematic build-up of such mesoscopic states photon by photon.

\subsubsection{Two-photon phase gate}
\label{sec: Two-photon phase gate}
In addition to such fundamental studies, the availability of a strong dispersive nonlinearity holds prospects for applications in quantum information processing. Of central interest here has been the implementation of deterministic gate operations, providing a complementary approach to probabilistic protocols based on linear optics \cite{Kok2007, Knill2001}. Rydberg EIT based photonic phase gates can be implemented in a number of ways, for instance, via the head-on collision of counter-propagating polaritons \cite{Friedler2005,Gorshkov2011} or by first storing two photons in collective spin wave states \cite{Barato2014} and using an additional microwave coupling to auxiliary Rydberg states to control the effective photon-photon interaction \cite{Maxwell2013, Maxwell2014}.

\begin{figure}[!b]
\begin{center}
\includegraphics[width=0.8\textwidth]{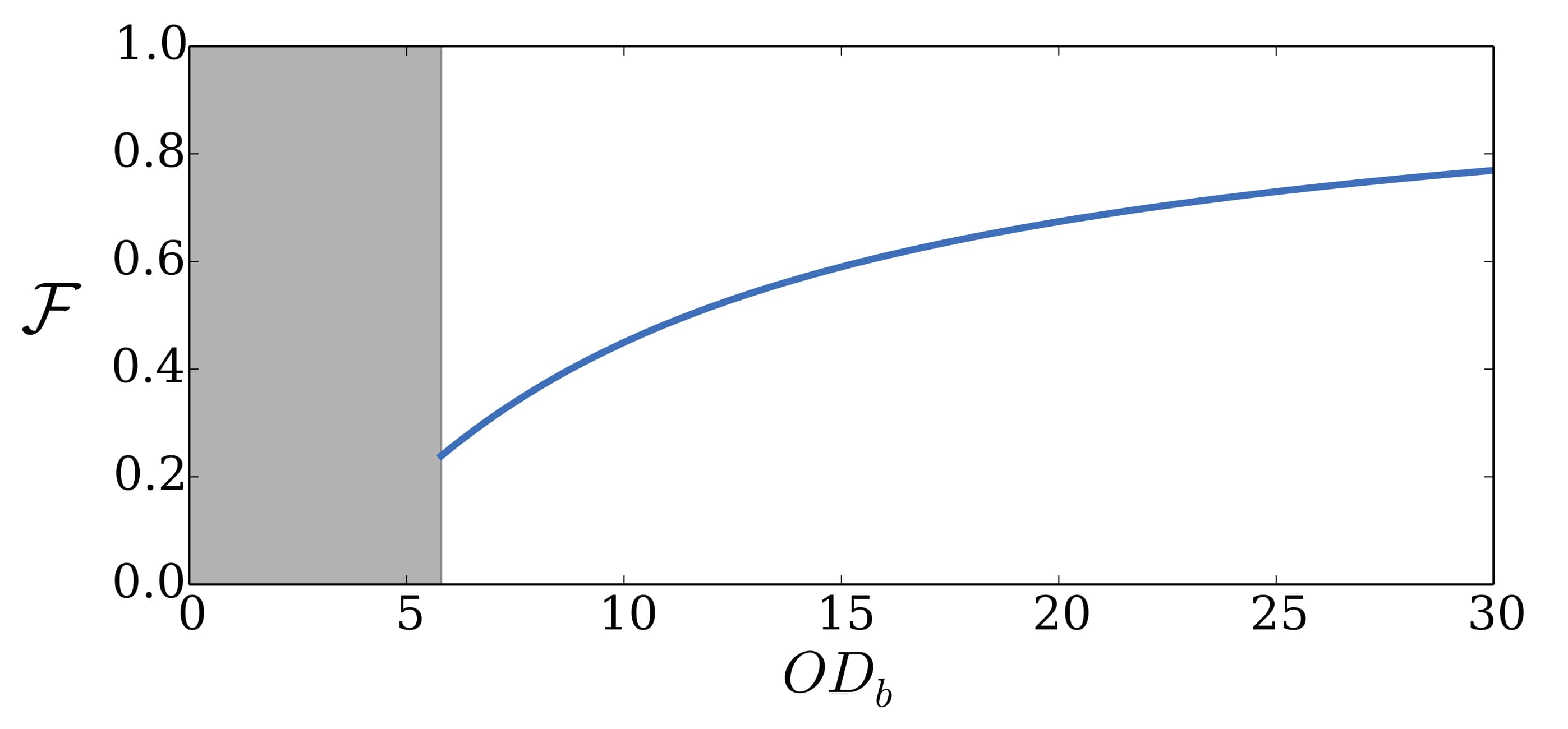}
\end{center}
\caption{\label{fig: CZ gate fidelity} Fidelity of a $\pi$-phase gate as a function of $OD_b$. The grey shaded region indicates the values of $OD_b$ over which a $\pi$ phase shift is impossible.}
\end{figure}

However, we will here focus on the conceptually simplest scheme which follows the same strategy as the optical-switching protocol described in Section \ref{sec: All-optical switch}, whereby one photon is made to propagate through another photon stored as a Rydberg spin wave. The photon propagation in this case is again governed by Eq.(\ref{propagation equation}) where the static susceptibility is now given by 
\begin{equation}
\chi(\tilde z, 0) \approx  - OD_b\left[ \left(\frac{\gamma}{\Delta}\right) \frac{1}{1 + \tilde{z}^6} - i \left(\frac{\gamma}{\Delta}\right)^2 \frac{1}{(1 + \tilde{z}^6)^{2}} \right],
\end{equation}
in the limit $\gamma\ll|\Delta|$. Accounting for the finite photon bandwidth to linear order in $\omega$, integration of Eq.(\ref{propagation equation}) gives the following simple solution for the output amplitude
\begin{equation}
\mathcal{E}(L, t) \approx e^{i\phi - \eta} \mathcal{E}(0, t - [L - r]/v_g)
\end{equation}
of the photon after having traversed the length $L$ of the medium, in terms of the incident photon field $\mathcal{E}(0, t)$. The Rydberg-blockade results in a phase shift $\phi$, an exponential amplitude attenuation of $e^{-\eta}$ and a modified transit time, as captured by the effective EIT medium length $L-r$, each of which are defined according to
\begin{align}
\label{phase}
& \phi = - OD_b  \left( \frac{\gamma}{\Delta} \right) \frac{2\pi}{3} \\
\label{attenuation}
& \eta = OD_b \left( \frac{\gamma}{\Delta} \right)^2 \frac{5\pi}{9} \\
\label{time delay}
& r = \left[ \frac{7 \pi}{9} - \frac{\Omega^2}{\Delta^2} \frac{5 \pi}{9} \right] z_b
\end{align}
The medium length modification can be either positive or negative, reflecting the fact that the group velocity, $g^2\rho c/\Delta^2$, of the effective two-level medium with the blockade region can be faster or slower than the underlying EIT group velocity $v_g=g^2\rho c/\Omega^2$ if $\Omega>|\Delta|$.

A high-fidelity gate operation within this setup thus requires a large single-photon detuning ($|\Delta| \gg \gamma$) in order to suppress absorption relative to nonlinear phase accumulation. Consequently, large values of $OD_b$ must again be achieved in order to compensate for the associated reduction to the photon-spinwave interaction. Since this scheme operates independently of the size $L$ of the medium, storage and retrieval can in principle be performed with arbitrarily high efficiencies. The maximum gate fidelity, $\mathcal{F}={\rm e}^{-2\eta}$, is hence limited only by scattering of the second photon. This is indeed confirmed by FIG.\ref{fig: CZ gate fidelity}, showing the expected fidelity for a $\pi$-phase gate as a function of $OD_b$. At large values of $OD_b$, it follows from Eq.(\ref{phase}) and Eq.(\ref{attenuation}) that the infidelity, $1-\mathcal{F}\approx (5\pi)/(2OD_b)$ exhibits a rather slow decrease with increasing $OD_b$. In addition, there is also a threshold value of $OD_b\sim6$ below which a phase shift of $\pi$ can not be reached. As experiments are currently limited to $OD_b\lesssim2$, this presents a serious limitation to practical implementations of gate operations. Recently, Tiarks et al. \cite{Tiarks2016} resolved this issue by working off two-photon resonance, i.e. outside the EIT-regime, and thereby succeeded to observe a large conditional phase shift of $\phi\approx\pi$.

\section{Summary and outlook}

Since its conception, the field of Rydberg-EIT has advanced at a rapid pace on both a theoretical and experimental level. The optical nonlinearities offered with this fundamentally distinct approach are the largest ever recorded and bestow unique opportunities for fundamental and applied science by manipulating freely propagating light at the ultimate quantum level. In this article, we have highlighted the many spectacular achievements already accomplished in this relatively young field of research, and attempted to give some insight into possible future directions. Nonetheless, there are several outstanding challenges and obstacles yet to be solved before many of these future perspectives can be brought to full fruition.

\begin{figure}[!t]
\begin{center}
\includegraphics[width=0.8\textwidth]{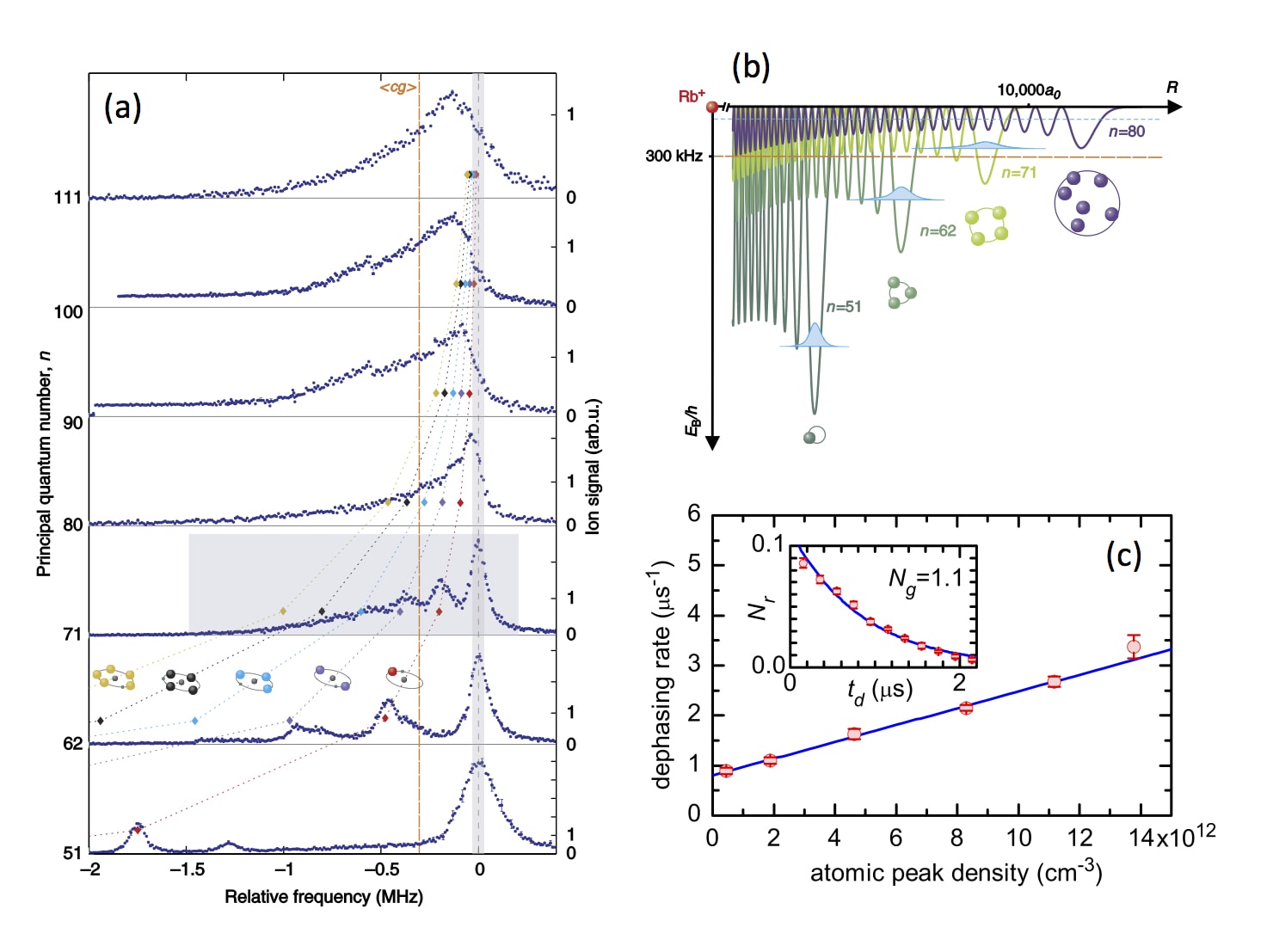}
\end{center}
\caption{\label{fig: Rydberg molecule} The low-energy scattering of a Rydberg electron and other ground state atoms placed within its wave function can lead to sizeable interactions shown in (b) for different principal quantum numbers of rubidium $nS_{1/2}$ states. (a) It can promote ultralong-range molecular states that appear as additional excitation resonances and eventually causes line broadening at larger $n$ or higher densities. (c) The latter degrades the efficiency of photon storage and retrieval in a Rydberg gas, with a loss rates that increases linearly with the atomic density. The inset shows the corresponding temporal decay of the retrieved photon number caused by this dephasing. Panels (a), (b) and (c) reproduced from \cite{Gaj2014} and \cite{Baur2014}}
\end{figure}

As stressed throughout this article, achieving high optical depths, $OD_b$, per blockade volume is the single most important requirement for accessing the deep quantum regime and implementing many of the described applications to high fidelity. One way to resolve this issue is to operate at higher atomic densities, which are readily available and have been demonstrated in numerous experiments on EIT with ground state atoms  \cite{Hau1999,Liu2001}. In the present situation, however, useful atom densities are limited by yet another type of interaction, namely that between Rydberg and ground state atoms. As first pointed out in \cite{Greene2000} the low-energy scattering between the Rydberg electron and a ground state atom can give rise to strong interatomic interactions (FIG.\ref{fig: Rydberg molecule}b) that can even promote the formation of exotic long-range molecules, as observed in a number of recent experiments \cite{Bendkowsky2009,Bendkowsky2010,Li2011}. As several ground state atoms start to penetrate the electronic Rydberg-state wave function, this ultimately causes inhomogeneous broadening of the otherwise narrow excitation line \cite{Balewski2013, Gaj2014} (FIG.\ref{fig: Rydberg molecule}a) and thereby spoils the quality of slow-light propagation and performance of light storage operations \cite{Baur2014} (FIG.\ref{fig: Rydberg molecule}c). 

A direct solution of this density-dependant dephasing could be achieved by controlling the underlying electron-atom scattering to suppress inhomogeneous broadening effects, or by using modified settings that can operate in the presence of static disorder. However, an alternative strategy to alleviating this issue might be to increase the effective optical depth of the medium by enhancing the single-atom coupling strength $g$, rather than by increasing the atomic density, e.g., by placing the Rydberg-medium inside an optical cavity. This would allow one to work with low density atomic media, so as to avoid the effects of electron-atom scattering, and yet maintain a strong light-matter coupling provided by the cavity enhancement. Such an approach has already been implemented in \cite{Parigi2012}, and holds significant promise for high-fidelity Rydberg-based photonic phase gates \cite{Das2016}, that might be further improved by employing heralded quantum gate strategies \cite{Borregaard2015}. 

From a different angle, the prospects for realising many-body systems of freely moving and strongly interacting photons provides further unique opportunities that are just beginning to be explored both in theory and experiment. In particular, the inherently driven-dissipative nature of such optical settings, in which photons continuously flow in and out of the system, leaves exciting open questions concerning their non-equilibrium physics, which currently are also being explored in a number of different photon-based systems  \cite{Sieberer2015}. However, the complicated nature of such driven-dissipative systems has restricted current experiments and theoretical treatments to the few-body regime.

On the experimental side, another major challenge emerges from the relatively large extent of the Rydberg blockade in relation to the achievable size of cold Rydberg gases. Similar to previous experiments on the Rydberg-excitation dynamics in such systems \cite{Schauss2012,Schauss2015}, this generally limits the number of simultaneously interacting polaritons to rather small values. Regarding the atomic dynamics, this obstacle was recently overcome by Zeiher et al. \cite{Zeiher2016} via Rydberg-dressing to lower lying states. In combination with a cavity-enhanced light-matter coupling, this strategy also appears to be a promising approach towards the regime of many interacting photons.

More generally, the concepts developed for cold Rydberg gases could tap into new potential for other physical settings, and may indeed open up new regimes of length and energy scales that are not accessible with cold Rydberg atoms. Indeed, recent work suggests that such diverse settings as thermal vapour cells \cite{Honer2011,Ripka2016}, semiconductor excitons \cite{Kazimierczuk2014} or nanophotonic structures \cite{Shahmoon2016,Douglas2016} provide a very positive outlook for future explorations. Surely, the range of exciting questions and perspectives will make Rydberg-EIT an active field of research for years to come. Considering the remarkable advance of the field over just the past five years, current challenges will be likely met in the near future and have the potential to prompt a paradigm shift in our approach to quantum nonlinear optics.

\section{Acknowledgments}
The advances reviewed in this article are the results from the combined efforts and innovative ideas of many colleagues over the past years. We express particular thanks to Charles Adams, Immanuel Bloch, Hans Peter B\"uchler, Stephan D\"urr, Ofer Firstenberg, Michael Fleischhauer, Alexey Gorshkov, Christian Gross, Sebastian Hofferberth, Matthew Jones, Thomas Killian, Igor Lesanovsky, Mikhail Lukin, Klaus M{\o}lmer, Tilman Pfau, Gerhard Rempe, Jeff Thompson, Vladan Vuleti\'c and Matthias Weidem\"uller for many inspiring discussions and collaborations. 

\bibliographystyle{apalike}
\bibliography{references}

\end{document}